\documentclass[%
 aip,
 amsmath,amssymb,
preprint,%
]{revtex4-1}
\usepackage{graphicx}  
\usepackage{dcolumn}   
\usepackage{bm}        
\usepackage{amssymb}   
\usepackage{amsmath}   
\usepackage{comment} 
\usepackage[caption=false]{subfig}
\usepackage{wrapfig}
\usepackage{float}
\usepackage{adjustbox}
\usepackage{multirow}
\usepackage{tabularx}
\usepackage{caption}

\newcolumntype{Y}{>{\centering\arraybackslash}X}

\usepackage{tikz}
\begin{document}
\preprint{AIP/123-QED}
\title{Exact solutions to planar emittance growth problems}
\author{B. Zerbe}\email{zerbe@msu.edu}
\author{P. M. Duxbury}\email{duxbury@msu.edu}

\affiliation{Department of Physics and Astronomy, Michigan State University}
\date{\today}
\begin{abstract}
This paper is the first in a series which develops the theory of emittance dynamics based on simple statistical reasoning.  
Emittance is a central quantity used to characterize the quality of electron microscopes, photon sources and particle beams.  
Emittance growth in high intensity charged particle beams is a particularly challenging 
non-equilibrium statistical physics problem in which effects such as disordered-induced 
heating and charge reorganization can lead to very rapid degradation of emittance and beam quality.  
The concepts of free energy and entropy have been utilized to improve conceptual understanding of 
emittance dynamics.   Here we develop a theory based on the second order cumulant of 
particle distributions and use this formulation to exactly solve several one dimensional problems.  
These solutions are important extensions of the existing results for the free expansion dynamics of pancake bunches 
used in ultrafast electron microscopy, which at short times are known to expand quadratically with time\cite{Siwick:2002_mean_field,Reed:2006_short_pulse_theory,Zerbe:2018_coulomb_dynamics}.   Here we show that the squared 
emittance of a strictly planar expanding bunch increases as a quadratic polynomial of time.    We compare theories 
based on individual particle trajectories with theories based on distributions and expand the foundations of theories 
based on individual particle trajectories, which we call the ``sample picture".   Our later work uses this formulation to 
derive generalized envelope equations which capture emittance growth effects in two and three dimensional systems.
\end{abstract}
\maketitle

\section{Introduction}
Non-equilibrium, and especially non-linear non-equilibrium, systems, 
are a central problem in modern theory.  A significant portion of modern theory 
focuses on the evolution of the phase-space distribution of particles,
and these approaches are generally theoretically grounded in Liouville's theorem
that states that the phase-space distribution function is constant along the
trajectory of the system.  The most central methods in near-equilibrium and
non-equilibrium statistical mechanics,
including linear response theory, Boltzmann Transport Theory, and
the Bogoliubov-Born-Green-Kirkwood-Yvon hierarchy, 
are based on such approaches, and numerous insights have been obtained from the
application of such models [see e.g. textbooks such as Huang, ``Statistical Mechanics"\cite{Huang:1987_book} 
or Kardar ``Statistical Physics of Particles"\cite{Kardar:2007_book}].    
Though the phase space distribution is often the central quantity, 
theories based on particle trajectories are also standard, 
particularly in computational approaches such as molecular dynamics simulations.    
Here we show that when the flow of a particle bunch is laminar, or close to laminar, 
it is possible to develop analytic theories for particle trajectories which can be used to 
develop theories for the dynamics of the emittance; as well as for the expansion energy 
and the energy spread of the beam.   This is important as high intensity beams of 
practical interest are close to laminar.

The discussion of particle beams is often based on continuous distribution functions in
$2D$, $6D$, or $6ND$ phase-space.
While the descriptions of the evolution of the 
full distribution function is a common goal
of traditional analytic techniques, specific problems often focus on understanding the evolution of specific
cumulants, e.g. means and covariances of these distributions.  
Furthermore the characteristic function, which can be expanded in terms of cumulants or modes,
can be thought of as an equivalent description
of the phase space distribution albeit in a Fourier-transformed space\cite{Papoulis:1991_book}. 
Thus the understanding of the moments or cumulants 
may provide adequate understanding of the distribution without having a full description.

Statistical emittance, henceforth called emittance, is  a statistical measure of the quality of an ensemble of particles.  It is central to 
the discussion of numerous bunch phenomena in the accelerator 
literature\cite{Reiser:1994_book,Buon:1992_phase_space,Luiten:2004_uniform_ellipsoidal,Reed:2009_imaging_review},
and optimizing the emittance is a goal of a large portion of the accelerator 
community\cite{Buon:1992_phase_space,Lawson:1992_emittance}.  
This measure is obtained from the second order
cumulants\cite{Lapostolle:1971_envelope_filamentation,Sacherer:1971_envelope}, and this formulation  
can be thought of
as the square root of the determinant of the covariance matrix\cite{Buon:1992_phase_space}.  In the statistics literature,
the determinant of the covariance matrix is sometimes referred to as the generalized variance, 
whose square root can be thought of as a generalized width (area, volume, etc.) in higher
dimensions\cite{Wilks:1932_gen_var}.  Therefore, emittance can be thought of as an operationalization of phase space
area (or volume if in $6D$), and accelerator physicists are specifically interested in the evolution of this
measure as well as the covariance measures from which it is derived.

The prevailing energy theory for emittance growth was principally developed in the 70's and 80's.
The connection between emittance growth and the field energy contained in an ensemble is largely attributed to 
Lapostolle\cite{Lapostolle:1971_envelope_filamentation} and was formalized 
under the assumption that the field felt at a specific location was identical to the mean-field at that 
location and that non-linearities in this field led to a free energy that could be 
thermalized within the beam\cite{Wangler:1985_emittance_relaxation,Hofmann:1986_emittance,Wangler:1991_emittance_growth}.  
Furthermore, Anderson described exact
solutions for emittance growth of some simple symmetric models in the presence of a 
focussing force\cite{Anderson:1987_emittance} of note analytically replicating the quarter plasma period
emittance explosion seen in simulation by Wangler et. al \cite{Wangler:1985_emittance_relaxation}. 
These ideas were extended
by Struckmeier to include Fokker-Planck stochastic effects\cite{Struckmeier:1994_fokker_planck,Struckmeier:2000_stochastic}.
These theories paved the way for models explaining beam halo formation\cite{Gluckstern:1994_analytic_halo,Wangler:1998_particle_core_review} and 
emittance compensation\cite{Serafini:1997_emittance_compensation,Rosenzweig:2006_emittance_compensation}.
We note that these theories generally assume continuous, cylindrical-symmetric beam conditions in the presence of a constant
focussing force.

Concurrent with the development of this energy theory of emittance growth, 
Lawson, Lapostolle, and Gluckstern argued that there is a close relation
between the entropy and emittance of a beam\cite{Lawson:1973_emittance}.  While this idea largely remained dormant
until the energy theory was worked out, it was reintroduced in the early 
90's\cite{Wangler:1991_emittance_growth,Lawson:1992_emittance} to address some issues not answered by the
energy theory, and this entropy interpretation gained some popularity, 
e.g. \cite{Struckmeier:1996_entropy,Brown:1996_entropy,Brown:1997_free_energy,OShea:1996_free_energy,OShea:1998_entropy,Boine:2015_intense_beams}.  However both the energy and the entropy theories have been criticized
as phenomenological as they lack any understanding of the mechanism or time dependence of emittance 
growth, and further that the application of thermodynamic language is largely counterproductive\cite{Bernal:2015_conceptual}.

Fundamental to all treatments, though, are the KV envelope equations
 initially derived through the identification of a very special 
$6D$ phase space distribution whose projection to any $2D$ phase space is uniform\cite{Kapchinskij:1959_KV}.
Sacherer presented a derivation of the KV envelope equations through the derivatives of the 
ensemble cumulants\cite{Sacherer:1971_envelope}.   A subtle definition in Sacherer's work was 
 a statistical definition of the envelope
whereas previous work\cite{Kapchinskij:1959_KV} and even some more recent work, e.g. \cite{Struckmeier:1994_fokker_planck},
use a more radial-like parameter for the envelope; we will maintain the statistical definition in this work. 
Moreover, these envelope equations can be shown to be equivalent to the more recent 
Analytic Gaussian (AG) equation derived by different means\cite{Michalik:2006_analytic_gaussian,Michalik:2009_AG_comparison,Berger:2010_AG_focus}.
Sacherer correctly argued that his 
derivation was more general than Kapchinsky and Vladimirsky's
derivation as it was an approximation of the bunch no matter the underlying phase-space distribution
and that the envelope equations predict the evolution of high intensity beams astoundingly well 
if the emittance dynamics is known \textit{a priori}\cite{Sacherer:1971_envelope}.    In some applications assumption of conserved emittance is a good approximation, however we are most interested in situations where this is not the case.  Below we use the term envelope equations to refer to either KV or AG equations.

While the envelope equations are used in essentially all theoretical investigations, their derivation from distribtuion theory is different than from the sample perspective.  It is not clear whether Sacherer understood this difference; specifically, the operator used in his definition of the statistics based on distribution theory enables the commutation of the average and the 
time derivative operations, which is only true in the distribution theory when collisions are neglected.  An important advantage of the sample perspective is that commutation of average and time derivative is exactly true.  In the field of statistics, the distinction between analyzing abstract, often continuous distributions and 
analyzing specific, discrete distributions
is distinguished by the terms population and sample statistics\cite{Lane:2017_intro_stats}, respectively.  
Using this vernacular, the standard Liouville theorem distribution-based approaches may be thought of
as a population theory whereas theories based on the statistics of individual, particle motion can be thought of as a
sample theory.  While both notations appear in the literature, they are generally treated as equivalent with no 
discussion of their differences.
We will call the development of a theory that emphasizes the statistical analysis of a single ensemble of
particles before applying any population-level insight the sample (statistics) perspective, we will refer to
all direct analysis of the phase space distribution as the population (statistic) perspective, and we will highlight
some of the theoretical differences that arise when adopting these different perspectives.  
We develop the sample perspective to garner insight into the mechanisms underlying emittance growth as
we argued above that such insight is largely absent from the population perspective.  We also note that the sample statistic approach leads naturally to analysis of finite size effects in non-equilibrium systems.  

The sample perspective benefits greatly from the Lagrangian description of fluid dynamics.  Previous work describing Coulomb
explosion\cite{Reed:2006_short_pulse_theory,Zerbe:2018_coulomb_dynamics,Zerbe:2019_relativistic_dynamics,Zerbe:2019_laminar_flow} 
has successfully employed the Lagrangian description to describe the density evolution from a population perspective.
Fundamental to these treatments is the identification of a map from a distribution particle's 
original position and momentum at some time to its position and momentum at some later time.  Within the sample perspective, 
if the statistics of this map are analyzed, we can obtain deterministic equations for the evolution of the emmitance and other measures of beam quality.
In fact, this is precisely the approach employed by Sacherer\cite{Sacherer:1971_envelope} 
before deriving his form of the envelope equations.  We will explore Sacherer's map later in this manuscript.

The paper is organized as follows.   In Section II, we first present the statistics of the sample perspective and contrast it with the population perspective.  While theses two perspectives are utilized as a matter of course in many areas of research such as stochastic  differential equations and statistical physics, the theoretical accelerator physics literature generally 
focuses on the population perspective; while here we focus on the sample perspective and note how it differs from the population perspective.  Section \ref{sec:map:gen} describes simple, exact, mathematical relations between sample statistics and key accelerator physics measures of beam quality, such as emittance, when a map between the initial phase coordinates and the phase coordinates at time $t$ can be written.  In Section \ref{sec:map:non} we show how these rules can be applied in the non-interacting case leading to an extremely simple interpretation on the importance of the emittance.   This analysis is general and may be applied to massless particles and to massive particle beams at any energy. In Section \ref{sec:map:1D} we employ sample theory to derive an exact equation for emittance growth in laminar planar models for charged particle beams where the force on each particle can be treated as a constant.  This case describes the exact evolution of the statistics of a planar symmetric laminar system under Coulomb repulsion, and by the term exact, we mean that given an exact measurement of the statistics at the initial time, the statistics such as emittance may be determined for all time.  We then validate this expression, show that it provides a good approximation even for some non-laminar cases, and further suggest modifications to the theory that should capture the non-laminar case exactly.  Up to that point, we will have discussed the exact solution of evolution of the second order statistics; however, the expected solution of the evolution of the statistics can also be obtained  from an initial particle distribution.   This case is typical of experiments, where the initial phase space distribution may be known, for example a Gaussian particle distribution in the case of photo-injectors used in many electron beam systems\cite{Bassetti:1980_Gaussian,Berger:2009_parameters,Gahlmann:2008_ultrashort,Robinson:2015_gun}. 
In Section \ref{sec:exp:gen} we show how to use arbitrary population distributions to obtain expectations for the emittance
growth using the results of Section \ref{sec:map:1D:gen}.  We then validate these expectations by drawing and evolving multiple $N$-particle ensembles from three different initial distributions: uniform (Section \ref{sec:exp:unif}), Gaussian (Section \ref{sec:exp:gauss}), and quadratic bimodal (Section \ref{sec:exp:bimod}).   These results show that the expectation of key statistics largely captures the emittance growth of the non-uniform distributions, and that the  remainder, or finite system variations, of the emittance growth in all cases is proportional to $\frac{1}{\sqrt{M}}$, where $M$ represents the number of macroparticles, suggesting that  its origin is stochastic.  In Section \ref{sec:KE}, we show how kinetic energy is naturally written in the sample perspective, and we show how the dynamics of the parameters introduced in previous sections can be used to understand the accounting of energy flow through various modes and how this  relates to emittance change.  We use this understanding to briefly examine disorder induced heating within the sample perspective. In Section \ref{sec:discussion} we return to the comparison between the population and sample perspectives, discuss some of the more subtle points, and emphasize caveats of using the sample perspective.  Finally, in \ref{sec:conclusion}, we summarize the main findings of the paper and propose further work that is needed to make this perspective relevant to many key unresolved issues in accelerator physics.

\section{The population and sample perspectives}\label{sec:perspectives}

We begin with the standard formulation of the non-equilibrium theory relevant to the accelerator physics field
with special attention to the role of collisions and the first and second order cumulants of the distribution.
Central to the theory is the continuous $6ND$ phase space distribution at time $t$, 
$f_{6ND}({\vec{x}}_1,{\vec{x}}_2,\cdots,{\vec{x}}_N,{\vec{p}}_1,{\vec{p}}_2,\cdots,{\vec{p}}_N;t)$.  
In $6ND$ space, Liouville's theorem applies and 
$f_{6ND}({\vec{x}}_1,{\vec{x}}_2,\cdots,{\vec{x}}_N,{\vec{p}}_1,{\vec{p}}_2,\cdots,{\vec{p}}_N;t) = f_{6ND}({\vec{x}}_1,{\vec{x}}_2,\cdots,{\vec{x}}_N,{\vec{p}}_1,{\vec{p}}_2,\cdots,{\vec{p}}_N;0)$
However, due to the complications with working with
$6ND$ space, most work confines itself to presentations of the more workable $6D$ distribution,
$f_{6D}(x,y,z,p_x,p_y,p_z;t)$, by introducing
a collision term
\begin{align}
  \frac{df_{6D}}{dt} = C(\vec{x};t)\label{eq:BTE}
\end{align}
for the time evolution of the distribution --- an equation commonly known as the 
Boltzmann transport equation (BTE).  If $C(\vec{x};t) = 0$ in the BTE --- the resulting equation is often called the 
Vlasov equation --- where the $6D$ phase space density function is conserved like its $6ND$ counterpart.  
Otherwise, a second function is introduced, $f_{6D,\Delta}(x,y,z,p_x,p_y,p_z;t)$, that accounts for the
collision term.  Namely
\begin{align}
  f_{6D}(x,y,z,p_x,p_y,p_z;t) = f_{6D}(x,y,z,p_x,p_y,p_z;0) \nonumber\\ 
                                                \quad\quad\quad+ f_{6D,\Delta(t)}(x,y,z,p_x,p_y,p_z;t)
\end{align}
where $ f_{6D}(x,y,z,p_x,p_y,p_z;t) $ has the normalization
constraint of $\int_{V_{6D}} f_{6D,\Delta}(x,y,z,p_x,p_y,p_z;t) ~ dV_{6D} = 0$
where $V_{6D}$ represents the entire $x,y,z,p_x,p_y,p_z$ space.
It should be obvious that if $C(\vec{x};t) = 0$ we can write $f_{6D,\Delta}(x,y,z,p_x,p_y,p_z;t) = 0$.

Furthermore, further reduction of the complexity of the phase space is often made by 
analyzing the marginal distribution integrated over two of the dimension, e.g.
\begin{align}
  f_{2D,x}(x,p_x;t) &= \int\int\int\int f_{6D}(\vec{x},\vec{p}) ~dp_z dp_y dz dy
\end{align}
and analogously for $y$ and $z$.
Assuming the phase space density is not correlated between dimensions,
we can write
\begin{align}
  f_{6D}&(x,y,z,p_x,p_y,p_z;t) =  \nonumber\\&\quad\quad f_{2D,x}(x,p_x;t) f_{2D,y}(y,p_y;t) f_{2D,z}(z,p_z;t) 
\end{align}

The expectation of some function of the random variables,
$a = a(x,y,z,p_x,p_y,p_z)$, is written as
\begin{align}
   \langle a\rangle &= \int_{V_{6D}} a f_{6D} ~dV_{6D}\label{eq:exp gen}
\end{align}
If $a_x = a_x(x,p_x)$ is only a function of $x$ and $p_x$ and there is no correlation between the 
dimensions in phase space we have
\begin{align}
   \langle a_x\rangle &= \int\int a_x f_{2D,x} ~dp_x dx\label{eq:exp}
\end{align}
which will be the general form of the expectation we examine to simplify the arguments.
If $a(x,y,z,p_x,p_y,p_z) = x^iy^jz^kp_x^lp_y^mp_z^n$ for $(i,j,k,l,m,n)$ in the non-negative integers,
$\langle a\rangle$ is known as a moment of order $i+j+k+l+m+n$.  The $0^{th}$ order moment is
generally set to $1$ and is usually called the normalization condition on $f_{6D}$.  Once normalized, all other 
moments are set by the details of the distribution.   Statistics are then functions of these moments,
and cumulants can be written in terms of these moments as well.  The first order moment and the first cumulant
are in fact the same thing.

Of special interest to physics are the covariance statistics.  Covariance statistics are defined by
\begin{align}
  \sigma_{a,b} = \langle a b\rangle  - \langle a\rangle \langle b\rangle \label{eq:def cov}
\end{align} 
If $a$ and $b$ are linear in $x,y,z,p_x,p_y,p_z$, these statistics are the second order cumulants.  Furthermore,
$\sigma_{a,a}$ is usually abbreviated as $\sigma_a^2$, known as the variance of $a$, so that the 
standard deviation $\sigma_a = \sqrt{\sigma_{a,a}}$ has a simpler notation.  In addition to the $\sigma$ notation,
we will also use $var(a)$ and $cov(a,b)$ for $\sigma_a^2$ and $\sigma_{a,b}$ when $a$ or $b$ are complicated.
Often
all relevant second order statistics are put in a $6 \times 6$ matrix, called the covariance matrix, whose
rows and columns can be thought of as labeled by $x,p_x,y,p_y,z,p_z$: 
\begin{align}
  \mathbf{COV}_{6D} = \begin{bmatrix}
                                     \sigma_x^2 & \sigma_{x,p_x} & \sigma_{x,y} & \sigma_{x,p_y} & \sigma_{x,z} & \sigma_{x,p_z}\\ 
                                     \sigma_{x,p_x} & \sigma_{p_x}^2 & \sigma_{y,p_x} & \sigma_{p_x,p_y} & \sigma_{z,p_x} & \sigma_{p_x,p_z}\\
                                     \sigma_{y,x} & \sigma_{y,p_x} & \sigma_y^2 & \sigma_{y,p_y} & \sigma_{y,z} & \sigma_{y,p_z}\\
                                     \sigma_{x,p_y} & \sigma_{p_x,p_y} & \sigma_{y,p_y} & \sigma_{p_y}^2 & \sigma_{z,p_y} & \sigma_{p_y,p_z}\\
                                     \sigma_{x,z} & \sigma_{z,p_x} & \sigma_{y,z} & \sigma_{z,p_y} & \sigma_z^2 & \sigma_{z,p_z}\\
                                     \sigma_{x,p_z} & \sigma_{p_x,p_z} & \sigma_{y,p_z} & \sigma_{p_y,p_z} & \sigma_{z,p_z} & \sigma_{p_z}^2 
                                     \end{bmatrix}
\end{align}
 Usually $\mathbf{COV}_{6D}$
is assumed to be block diagonal with elements, $\mathbf{COV}_{x}$, $\mathbf{COV}_{y}$, $\mathbf{COV}_{z}$ where
\begin{align}
  \mathbf{COV}_{x} = \begin{bmatrix}
                                     \sigma_x^2 & \sigma_{x,p_x}\\ 
                                     \sigma_{x,p_x} & \sigma_{p_x}^2 
                                   \end{bmatrix}
\end{align}
and analogously for $y$ and $z$.  Of interest to the accelerator physicist, the $2D$ emittance is defined by
\begin{align}
  \epsilon_{x,p_x}^2 &= \frac{1}{m^2c^2}\lvert \mathbf{COV_x} \rvert\nonumber\\
							&= \frac{1}{m^2c^2} cov(x,p_x) \nonumber\\
                                &= \frac{1}{m^2c^2}(\sigma_x^2 \sigma_{p_x}^2 - \sigma_{x,p_x}^2)\label{eq:def em}
\end{align}
where $\lvert\cdot\rvert$ indicates the determinant of the contained matrix.  As 
$\sqrt{\lvert \mathbf{COV_x} \rvert}$ can be thought of as an area of $f_{2D,x}$ in $x$-$p_x$ space,
emittance is often visualized as the area of an ellipse with the appropriate axis lengths determined by the statistics.

The sample perspective is very similar to the population perspective except instead of the continuous $6D$ phase
space function, $f_{6D}$, being central to the definition of the statistics, the discrete $N$ particle sample itself takes this role.
Specifically, the mean operator
\begin{align}
  \bar{a} = \frac{1}{N} \sum_{i=1}^N a_i\label{eq:mean}
\end{align} 
where $a_i$ is the $i^{th}$ value of the random variable $a$ replaces the role of the expectation operator, $\langle \cdot \rangle$. 
Often $\bar{a}$ is used to estimate the value of $\langle a \rangle$; 
specifically, it is fairly straightforward to prove
\begin{align}
  \langle \bar{a} \rangle = \langle a \rangle
\end{align}
although the exact distribution of $a$ may be of interest to statisticians.
However, we emphasize that these population and sample averages are distinct and care
needs to be taken when applying equations meant for one to the other.

To keep these operators and the statistics derived from them distinct, 
we introduce new notation (borrowed from the field of statistics) for the
sample covariances.  Specifically, 
\begin{align}
  s_{a,b} &= \overline{ab} - \bar{a} \bar{b}\label{eq:sample cov}
\end{align}
where again $a$ and $b$ are random variables.  As is the case for the moments,
it is fairly straightforward to show
\begin{align}
  \langle s_{a,b} \rangle = \sigma_{a,b}
\end{align}
However, this is not the case for the sample emittance defined by
\begin{align}
  \varepsilon_{x,p_x}^2 &= \frac{1}{m^2c^2} (s_x^2 s_{p_x}^2 - s_{x,p_x}^2)\label{eq:sample em}
\end{align}
Namely, the expected emittance is
\begin{align}
  \langle \varepsilon_{x,p_x}^2 \rangle &= \frac{1}{m^2 c^2} (\langle s_x^2 s_{p_x}^2\rangle - \langle s_{x,p_x}^2 \rangle)\nonumber\\
                                                             &= \frac{1}{m^2 c^2} (cov(s_x^2,s_{p_x}^2) + \langle s_x^2\rangle \langle s_{p_x}^2\rangle\nonumber\\
                                                             &\quad+ \langle s_x^2\rangle \langle s_{p_x}^2\rangle - var(s_{x,p_x}) - \langle s_{x,p_x} \rangle^2)\nonumber\\
                                                             &= \frac{1}{m^2 c^2} (\sigma_x^2 \sigma_{p_x}^2 -  \sigma_{x,p_x}^2 \nonumber\\
                                                             &\quad + cov(s_x^2,s_{p_x}^2)  - var(s_{x,p_x}))\nonumber\\
                                                             &=\epsilon_{x,p_x}^2 + \frac{1}{m^2 c^2} ( cov(s_x^2,s_{p_x}^2)  - var(s_{x,p_x}))\label{eq:em relation}
\end{align}
which differs from the population emittance by $\frac{1}{m^2 c^2} ( cov(s_x^2,s_{p_x}^2)  - var(s_{x,p_x}))$.  

In addition to this difference in emittance, there is a more fundamental difference between the sample and the population
perspectives that arises in the derivative of the operators, 
which can be thought of as a map relating the
statistics at differentially different times. 
We first derive the differential form in the sample perspective.   It is obvious that 
the time derivative of the mean of a random variable (assuming $N$ is held constant) is,
\begin{align}
  \frac{d}{dt} \bar{a} &= \overline{\dot{a}}
\end{align}
where $\dot{a}$ is shorthand for $\frac{da}{dt}$. This
results in the covariance statistics time derivatives being
\begin{align}
  \frac{d}{dt} s_{a,b} &= s_{\dot{a},b} + s_{a,\dot{b}}.\label{eq:dsdt}
\end{align}
This is the map Sacherer used in his derivation of the envelope equations\cite{Sacherer:1971_envelope}.    
Here we call Eq. (\ref{eq:dsdt}) applied to the phase space coordinates the statistical kinematics.  Specifically,
for the 2D phase space coordinates in the $x$-direction, the statistical kinematics are
\begin{subequations}
\begin{align}
  \frac{d}{dt} s_{x}^2 &= \frac{2}{m} s_{x,p_x}\label{eq:dsx^2dt}\\
  \frac{d}{dt} s_{x,p_x} &= \frac{1}{m} s_{p_x}^2  + s_{x,F_x}\label{eq:dsxpxdt}\\
  \frac{d}{dt} s_{p_x}^2 &= 2 s_{p_x,F_x}\label{eq:dspx^2dt}
\end{align}
\end{subequations}
where $F_x = \frac{dp_x}{dt}$.  
These equations are exact just as the typical kinematic equations from introductory physics are exact.  
Furthermore, the time derivate of the sample emittance can be written as
\begin{align}
  \frac{d \epsilon_{x,p_x}^2}{dt} &= \frac{2}{m^2c^2}\left( s_x^2 s_{p_x,F_x} - s_{x,p_x}s_{x,F_x} \right)\label{eq:demdt}
\end{align}
where this equation is again exact.

These statistical kinematic equations can be used to obtain a 
$2^{nd}$ order ODE that is the non-relativistic form of Sacherer's envelope equations,
 by noting that $\frac{d}{dt} s_{x}^2 = 2 s_x \frac{ds_x}{dt}$
where $s_x = \sqrt{s_{x,x}}$.  
Combining this
observation with Eq. (\ref{eq:dsx^2dt}) gives
\begin{align}
  \frac{ds_x}{dt} = \frac{1}{m} \frac{s_{x,p_x}}{s_x},\label{eq:dsxdt}
\end{align} 
which we will use later.
Then taking the derivative of this equation and using Eq. (\ref{eq:dsxpxdt})
gives 
\begin{align}
  \frac{d^2s_x}{dt^2} = \frac{1}{m} \frac{s_{x,F_x}}{s_x} + \frac{c^2 \varepsilon_{x,p_x}^2}{s_x^3},\label{eq:dsqsxdtsq gen}
\end{align}
Again, this $2^{nd}$ order ODE is exact meaning that no assumptions or approximations have been made.  This equation together
with Eq. (\ref{eq:demdt}) represents the same degrees of freedom expressed in our statistical kinematics equations (or equivalently
the Corant Snyder parameters\cite{Buon:1992_phase_space}).
The envelope equations can then by obtained by making a single assumption --- that $F_x$ is the linear force seen in a
uniform ellipsoid, i.e $F_x = \frac{ 3 N e^2 }{40 \sqrt{5} \pi \epsilon_0 s_x^3} \alpha\left(\frac{s_y}{s_x},\frac{s_z}{s_x}\right)  x$
where $\alpha(a,b) = \int_0^{\infty} \frac{1}{(1+\lambda)^{3/2}\sqrt{a^2+\lambda}\sqrt{b^2+\lambda}} d\lambda$
giving
\begin{align}
  \frac{d^2s_x}{dt^2} =  \frac{ 3 N e^2 }{40 \sqrt{5} \pi m \epsilon_0 s_x^2} \alpha\left(\frac{s_y}{s_x},\frac{s_z}{s_x}\right) + \frac{c^2 \varepsilon_{x,p_x}^2}{s_x^3},\label{eq:dsqsxdtsq}
\end{align}
which is the non-relativistic elliptical envelope equation, which we discuss further in future work.
Note that this assumption results in Eq. (\ref{eq:demdt}) being $0$, which means emittance
is conserved; in fact, any assumption of emittance conservation is essentially 
equivalent to an assumption of the self-force inside the bunch
being linear.  
This means that Sacherer's envelope models\cite{Sacherer:1971_envelope}
assume such linear forces for a number of non-uniform distributions at least in the approximate sense.
Eq. (\ref{eq:demdt}) can be ignored and the entire evolution of the statistics can be described by Eq. (\ref{eq:dsqsxdtsq}) under
such an assumption.  

An analogous equation for the time derivative of the
population statistics can only be derived if the collision term, $C(\vec{x};t)$ in Eq. (\ref{eq:BTE}), is assumed to be
zero.  Under such an assumption, we can introduce a map, $\vec{\alpha}(x,y,z,p_x,p_y,p_z,t)$, 
that is explicitly dependent on time 
and that relates the Langrangian particle's phase space coordinates at time $t$ to its phase space coordinates at time $0$.  Another
way to state Liouville's theorem is that the phase space is incompressible, so this transformation has a Jacobian of $1$.
As a result, the expectation of the random variable may be written as
\begin{align}
  \langle a_x \rangle =  \int\int& a_x(x,p_x,t) f_{6D,x}(x,p_x;0) ~dp_x dx
\end{align}
Now $f_{2D,x}(x,y,z,p_x,p_y,p_z;0)$ is no longer dependent on time, but $a_x(x,p_x,t)$ is --- that is we have
moved the time dependence from $f_{2D,x}$ to the random variable.  Thus
\begin{align}
  \frac{d}{dt}  \langle a \rangle &= \left\langle \frac{d a}{d\vec{\alpha}} \frac{d\vec{\alpha}}{dt} \right\rangle
\end{align}
Unfortunately, this looks a lot more complicated than it really is --- as the statistical reasoning using the population
perspective can become fairly abstract and complex.  To simplify,
a concrete example of this is that
\begin{align}
  \frac{d}{dt} \langle x \rangle &= \frac{1}{m} \langle p_x \rangle
\end{align}
in the non-relativistic regime.
Furthermore, the covariance between $a$ and $b$ for $a,b$ in ${x,y,z,p_x,p_y,p_z}$ becomes
\begin{align}
  \frac{d}{dt} \sigma_{a,b} &= \sigma_{\dot{a},b} + \sigma_{a,\dot{b}}
\end{align}
analogous to Eq. (\ref{eq:dsdt}).  Again, though, this is only under the assumption of no collisions, and when collisions
are present, this mathematical reasoning fails and generally 
\begin{align}
  \frac{d}{dt} \sigma_{a,b} = \sigma_{\dot{a},b} + \sigma_{a,\dot{b}} + C_{a,b}(t)
\end{align}
where $C_{a,b}$ accounts for collisions.  In other words, Eq. (\ref{eq:dsdt}) can only be
derived in the population perspective if collisions are ignored or if an additional
term is added.  We note that Kapchinsky and Vladimirsky made the explicit assumption that
their $6D$ phase space volume was conserved meaning that their model was collision-less.  
It is reasonable to think that Sacherer may have implicitly made the same assumption
as he continued to use the population statistics notation; on the other hand, it is equally reasonable to believe that Sacherer thought of the problem from the sample perspective.

In conclusion, there are at least these two differences between the sample and population perspectives: 1.
a difference in the emittance and 2. a difference in the time derivatives of the moments.  
While the statistical kinematics 
are exact in the sample perspective, they are only exact in the population perspective under the assumption
of a collision-less (Vlasov) model or if an exact collision term is available and is included in the kinematics. 
Furthermore, as demonstrated below, the sample perspective lends itself to some very simple, convincing 
analyses whose corresponding analyses in the population
perspective are intricate.  This is not to say the population perspective is in anyway
inferior to the sample perspective, just different.  For the rest of the paper we focus on the sample perspective
and the analysis of the dynamics of second order sample statistics and specifically the emittance dynamics.

\section{Exact second order sample statistics evolution}\label{sec:map}

\subsection{General considerations}\label{sec:map:gen}
We return to the definition of the average of a physical parameter, $a$,
across all particles in an ensemble, Eq. (\ref{eq:mean}).  In the language of statistics,
$a$ is called a random variable and $a_i$ is the value of the random variable for the $i^{th}$ particle
within a specific sample.
As classical mechanics is deterministic,  $a_i$ can be mapped back to the values of the initial phase coordinates
for that particle, ${\vec{x}}_{0,i}$ and ${\vec{p}}_{0,i}$.   This mapping can be used effectively in the case of laminar flow, 
however for chaotic systems a mapping of this type for each particle is usually not analytically tractable.    
Here we focus on near laminar flow.
In most particle beams, parameters like time or mass are the same for all particles. 
Statisticians call parameters that are the same for all particles ``scalars", but
we will call such parameters ensemble constants to avoid confusion with the standard meaning of ``scalars" in physics. 
When the maps between the initial phase position
and the phase position at some time $t$ are analytically intractable, molecular dynamics can be thought
as a means to provide such a map.  However, when the flow is near laminar this map can sometimes be constructed analytically without reference to simulation techniques. This can always be achieved in the differential form, which we call the statistical kinematics. Using such maps in  Eqs. (\ref{eq:sample cov}) and (\ref{eq:sample em})
allows us to obtain analytic forms for the evolution of the statistics, such as emittance.  To carry out this analysis we use the easily proved statistical relation
\begin{align}
  s_{Wa+Xb,Yc + Zd} &= WY s_{a,c} + WZ s_{a,d} + XY s_{b,c} + XZ s_{b,d}\label{eq:cov sum}
\end{align}
where the quantities $a,b,c, d$ are random variables (usually position or momentum
of a particle) and $W, X, Y, Z$ are ensemble constants  (for instance the mass or time).

Notice, that nowhere in this discussion of the evolution of the statistics have we mentioned the
underlying phase-space distribution of the particles, just the value of the random variables.  
This is due to the fact that this description 
is independent of the phase-space distribution unless the map itself depends on it.

Before we get to concrete examples, we briefly discuss decomposing the momentum within the sample
perspective.    
Specifically, let $ {\bar{p}}_{x_i} $ represent the expected $x$-momentum at $x_i$, namely
\begin{align}
  p_{i} &= {\bar{p}}_{x_i} + \delta_i 
\end{align}
In other words $\delta_i$ is the deviation of the $i^{th}$ $x$-momentum from
the expected value.  In this work we will follow the standard practice in accelerator physics, which is to assume that the 
momentum fluctuation $\delta_i$ is independent of $x$.   Our next question is, ``What do we use for the expected value of the momentum?"  The answer
to this question depends on the system, and we focus on systems where there is a linear relation between the 
average momention and the average position.   This is the standard convention in experiments where systems with this linear relation can be precisely controlled using electromagnetic lens' and RF cavities. Experimentalists extract this linear relation from their data by choosing the line of best fit for the distribution in 
$x$-$p_x$ space.   Due to this linear relation, we can write the expectation for the momentum at $x_i$ as,  
\begin{align}
 {\bar{p}}_{x_i} &= {\bar{p}}_{x} + \frac{s_{x,p_x}}{s_x^2}(x_i - \bar{x})
\end{align}
where the quantity $\frac{s_{x,p_x}}{s_x^2}$ is the slope of the line of best fit of the $x-p_x$ phase space data
and ${\bar{p}}_{x}$ represents the average $x$-momentum of the entire ensemble.
Under these definitions, we can write the momentum variance as
\begin{align}
  s_{p_x}^2 &= \frac{1}{N} \sum_{i=0}^N (p_{x_i} - {\bar{p}}_x)^2 \nonumber\\
                   &= \frac{1}{N} \sum_{i=0}^N \left(\frac{s_{x,p_x}}{s_x^2}(x_i - \bar{x}) + \delta_i\right)^2\nonumber\\
                   &= \frac{s_{x,p_x}^2}{s_x^2} + 2 \frac{s_{x,p_x}}{s_x^2} s_{x,\delta} + s_{\delta}^2\label{eq:lin mom width}
\end{align}
As we assume that $\delta_i$ is independent of $x_i$, then $s_{x,\delta} = 0$ and the second term drops out.
Now we introduce the important quantity, 
\begin{align}
  \eta_x = s_\delta = \sqrt{s_{p_x}^2 - \frac{s_{x,p_x}^2}{s_x^2}} = \frac{mc \varepsilon_{x,p_x}}{s_x}\label{eq:etax}
\end{align}
We call $\eta_x$ the
local momentum spread as it has units of momentum and represents the average width of the momentum about
the expected position-momentum line.  This quantity already, in its squared form, appears in the literature in many works and under 
different names\cite{Maxson:2013_DIH,Michalik:2006_analytic_gaussian,Michalik:2009_AG_comparison,Berger:2010_AG_focus}, often with emphasis on the emittance relation (last relation in Eq. 
(\ref{eq:etax}).
We introduce a second important statistical parameter which also has units of momentum,
\begin{equation}
\mu_x = \frac{s_{x,p_x}}{s_x}
\end{equation}
Notice that $\mu_x$ can be written as $\mu_x = m \frac{d s_x}{dt}$ by Eq. (\ref{eq:dsxdt})-- that is
$\mu_x$ can be though of as the momentum involved with the changing of the $x$-standard deviation of the 
ensemble.
We call $\mu_x$ the linear flow momentum as it is the flow that is often attributed to expansion or contraction of an ensemble
as calculated using the line of best fit.
The local momentum spread may be written in terms of linear flow momentum as;
\begin{equation}
\eta_x^2 = s_{p_x}^2 - \mu_x^2
\end{equation}
Finally $\frac{1}{2m} \eta_x^2$ and $\frac{1}{2m} \mu_x^2$ have units of energy and 
represent the kinetic energy stored in the local momentum spread and in the linear flow momentum;
we will call these quantity the linear heat along $x$ and the linear flow energy along $x$, respectively.

While this discussion so far has been fairly abstract, we demonstrate two concrete cases where 
the particle trajectories and hence the maps required for the analysis can be analyticaly calculated:
 1. the non-interacting case with no external forces and 
2. the $1D$, planar-symmetric model.  We will show how the above insights can be used to predict 
both statistics evolution as well as to classify effects that drive emittance growth.  We note here
that these two cases are not the only two situations that can employ this mapping approach; for instance,
this mapping approach can be applied to cases where the interaction of the particles is replaced by some mean-field force
that lends itself to calculation
and hence the trajectory of each particle can be deduced.  A second example is non-interacting particles that are influenced
by external fields.  As such situations arise in many accelerator physics 
applications, the statistical approach described here is quite general.

\subsection{Non-interacting, freely-expanding, relativistic particles}\label{sec:map:non}

Consider non-interacting particles with no force on them.  The position and momentum of such 
a non-interacting particle with energy $E$ are
\begin{subequations}
  \begin{align}
    \vec{x} &= \vec{x_0} + \frac{c^2}{E} {\vec{p}}_0 t\label{eq:non-int pos} \\
    \vec{p} &= {\vec{p}}_0\label{eq:non-int mom}
  \end{align}
\end{subequations}
Notice that if the particle has a non-zero mass, $\frac{c^2}{E} = \frac{1}{\gamma m}$ and the
position equation reduces to the standard equation 
$\vec{x} = \vec{x_0} + \frac{1}{\gamma m} {\vec{p}}_0 t $; however, Eqs. (\ref{eq:non-int pos}) 
and (\ref{eq:non-int mom}) also
apply to mass-less particles like photons.  
We consider the statistics in the $x$-direction as the statistics in the other directions are analogous; 
Eqs. (\ref{eq:non-int pos}) 
and (\ref{eq:non-int mom}) in the $x$-direction are
\begin{subequations}
  \begin{align}
    x &= x_0 + \frac{c^2}{E} p_{0,x} t\label{eq:non-int pos:x} \\ 
    p_x &= p_{0,x}\label{eq:non-int mom:x}
  \end{align}
\end{subequations}
Notice that $x_0$, $p_{0,x}$, and $E$ are random variables whereas $c$, and $t$ are ensemble constants.  Using 
Eq. (\ref{eq:cov sum}), we obtain 
\begin{subequations}
  \begin{align}
    s_{x}^2 &= s_{x_0 + \frac{c^2}{E} p_{0,x} t}^2\nonumber\\
                          &= s_{x_0}^2 + c^4 t^2 s_{\frac{p_{0,x}}{E}}^2 + 2c^2 t s_{x_0,\frac{p_{0,x}}{E}}\label{eq:spa width}\\
    s_{x,p_x} &= s_{x_0 + \frac{c^2}{E} p_{0,x} t, p_{0,x}}\nonumber\\
                           &= s_{x_0,p_{0,x}} + c^2 t s_{p_{0,x},\frac{p_{0,x}}{E}} \label{eq:spa-mom cov}\\
    s_{p_x}^2 &= s_{p_{0,x}}^2 \label{eq:mom width}
  \end{align}
\end{subequations}
From these relations, we find an explicit expression for the emittance evolution:
\begin{align}
    \varepsilon_{x,p_x}^2 &= \varepsilon_{x_0,p_{0,x}}^2 + \frac{c^2}{m^2} t^2 \left(s_{p_{0,x}}^2 s_{\frac{p_{0,x}}{E}}^2 - s_{p_{0,x},\frac{p_{0,x}}{E}}^2\right) \nonumber\\
                                       &\quad\quad \frac{1}{m^2} t  \left(s_{p_{0,x}}^2 s_{x_0,\frac{p_{0,x}}{E}} - s_{x_0,p_{0,x}}s_{p_{0,x},\frac{p_{0,x}}{E}}^2\right)\label{eq:non em gen}
\end{align}

In most treatments of ensembles, the energy is assumed to be approximately the same for all particles\cite{Reiser:1994_book}; that is, $E$ is treated like an ensemble
constant instead of a random variable.  In this case, the $E$ can be pulled out of the expressions, and the statistics may be written as
\begin{subequations}
  \begin{align}
    s_{x}^2 &= s_{x_0}^2 + \frac{c^4 t^2}{E^2} s_{p_{0,x}}^2 + 2 \frac{c^2 t}{E} s_{x_0,p_{0,x}}\label{eq:spa width:E const}\\
    s_{x,p_x} &= s_{x_0,p_{0,x}} + \frac{c^2 t}{E} s_{p_{0,x}}^2 \label{eq:spa-mom cov:E const}\\
    s_{p_x}^2 &= s_{p_{0,x}}^2 \label{eq:mom width:E const}
  \end{align}
\end{subequations}
Further under this $E$ is an ensemble constant assumption, 
note that the emittance is conserved, i.e. $\varepsilon_{x,p_x}^2 = \varepsilon_{x_0,p_{0,x}}^2$.
With this observation, we can determine when the width of the distribution reaches
a minimum by taking the time derivative of Eq. (\ref{eq:spa width}) and setting the derivative to zero
and then solving for the time
\begin{align}
  t_{min~width} &= - \frac{E}{c^2} \frac{s_{x_0,p_{0,x}}}{s_{p_{0,x}}^2}
\end{align}
Plugging $t_{min~width}$ into Eq. (\ref{eq:spa width}), we obtain the minimum width of
the distribution to be
\begin{align}
  s_{x}(t_{min~width}) &= \frac{E}{c} \frac{\varepsilon_{x,p_x}}{s_{p_x}}\label{eq:min width}
\end{align}
This means that the spatial width of a non-interacting distribution at the focal point
is determined by the size of the emittance and the width of the distribution in momentum space. 
According to Eq. (\ref{eq:lin mom width}), $s_{p_x}$ can be controlled by manipulating $|s_{x,p_x}|$ 
when $s_x$ and the local momentum spread are held fixed.  This is effectively how a thin 
lens work.  Thus, the minimum width of an ensemble is only limited by the ability to adjust
$|s_{x,p_x}|$ --- and there are no other effects besides controlling when and how small a bunch focusses. 
This property of the minimum spatial width being set by the emittance when given a specific focussing $|s_{x,p_x}|$
is one of the main reasons that emittance is of interest. 
  
While this formulation is fairly trivial, we have not seen this approach in the literature and it leads to both new results for simple models and a framework for interacting systems.    
For example,  we point out that the
spatial-momentum covariance at $t=t_{min~width}$ is
\begin{align}
  s_{x,p_x}(t_{min~width}) &= 0\label{eq:s_xpx zero}
\end{align}
for this non-interacting case.
It is fairly trivial to show that this covariance is always zero at the minimum width
even in the presence of interactions as it is simply a result of the kinematics; which follows from the general expression,
\begin{align}
  \frac{ds_x}{dt} = \frac{s_{x,p_x}}{s_x}
\end{align} 
As the focal point is an extremum of $s_x$, it is then obvious that Eq. (\ref{eq:s_xpx zero})
is true in general.

\subsection{One dimensional models with Coulomb forces}\label{sec:map:1D}
\subsubsection{General 1D emittance phase-space dynamics}\label{sec:map:1D:gen}

Consider an ensemble of $N$-particles each with mass $m$ within the $1D$, planar model. 
This model and models equivalent to it have been used extensively to study the spreading dynamics of 
pancake bunches generated by ultrafast 
photo-injectors\cite{Siwick:2002_mean_field,Reed:2006_short_pulse_theory,Zerbe:2018_coulomb_dynamics}.  
Here we precisely resolve the emittance dynamics within these models. 

\begin{figure*}
  \centering
  \begin{tabular}{ccc}
   \subfloat[$v_0 = 0$ ]{\includegraphics[width=0.3\textwidth]{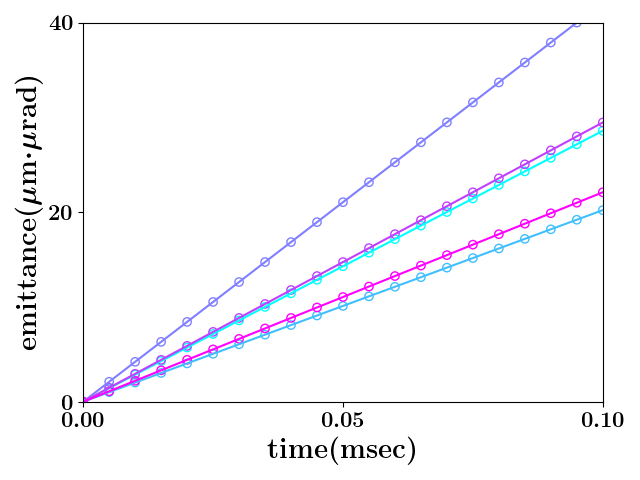}}&
    \subfloat[$v_0 = 4\times10^4 \frac{1}{s} x_0$ ]{\includegraphics[width=0.3\textwidth]{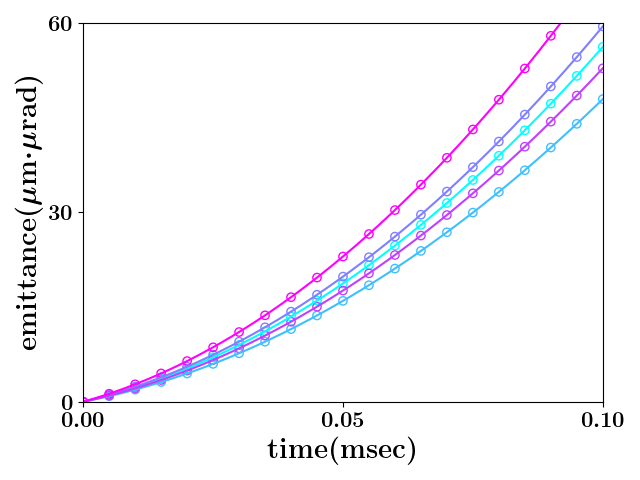}}&
    \subfloat[$v_0 = -4\times10^4 \frac{1}{s} x_0$ ]{\includegraphics[width=0.3\textwidth]{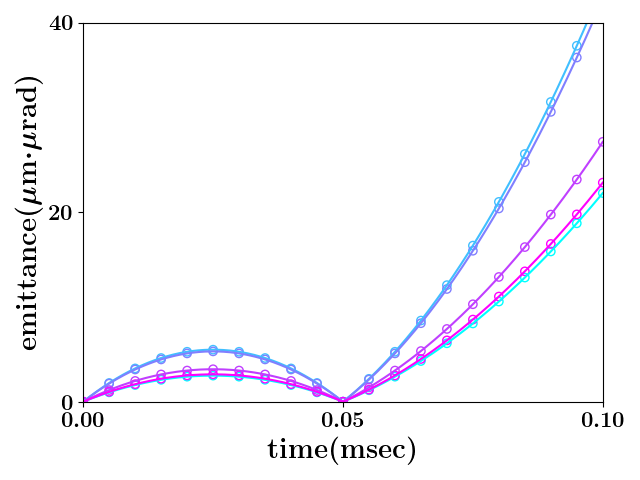}}  \end{tabular}
  \caption{\label{fig:emittance growth}
  Graphs of emittance growth under planar symmetry for ensembles of $1,000$ planar particles with 
  $\Sigma_{tot} = 1.6 \times 10^{-15} ~ \frac{C}{m^2}$ drawn from the uniform spatial distribution.
   The initial velocity chirp labels each graph.  Different colors represent five different samples of the 
   $1,000$ planar particles.
  Solid lines represent the theoretical prediction based on the initial statistics, 
  as detailed in Eq. (\ref{eq:time dependent determinant}), and open circles represent simulated results.
  Notice the exact agreement between the theory and simulation. 
  The curious bump in the last figure is a space-charge effect resulting from
  the fact that the particles come to rest and reverse their direction at nearly the same time ---
  this bounce-back situation causes the momentum spread to drop precipitously leading to
  a dramatic decrease in the emittance.}
\end{figure*}

Label the positions, velocities, and accelerations of the $i^{th}$ particle in the ensemble by 
$x_i$, $v_i$, and $a_i$, respectively.  Assuming that the particles in the bunch obey laminar flow, $a_i$ is 
a constant and the trajectory of each particle is described by
\begin{subequations}
\begin{align}
  x_i &= x_{i,0} + v_{i,0} t + \frac{1}{2} a_i t^2 \label{eq:1D kin:x}\\
  v_i &= v_{i,0} + a_i t\label{eq:1D kin:v}
\end{align}
\end{subequations}
where again $0$ in the subscript indicates the initial value of the parameter.  
Notice that Eqs. (\ref{eq:1D kin:x}) and (\ref{eq:1D kin:v}) are again
maps between the phase space at time $t$ and the initial phase space, and therefore
the approach we used for the non-interacting case can be used here as well.  Substituting
Eqs. (\ref{eq:1D kin:x}) and (\ref{eq:1D kin:v}) into Eq. (\ref{eq:sample em}), we 
obtain
\begin{align}
  \varepsilon_{x,p_x}^2 &= \frac{1}{c^2}|\mathbf{A}(t)|\label{eq:time dependent determinant}
\end{align}
where $|\cdot|$ again represents the determinant of the contained matrix and
$\mathbf{A}(t)$ is the $4\times4$ matrix
\begin{align}
  \mathbf{A}(t) &=  \left(\begin{array}{cccc}
                            0                    & \frac{1}{2} t^2 & -t               & 1\\
                            \frac{1}{2} t^2 & s_{x_0}^2         & s_{x_0,v_{0}} & s_{x_0,a}\\
                            -t                    & s_{x_0,v_{0}}      & s_{v_{0}}^2    & s_{v_{0},a}\\
                            1                    & s_{x_0,a}      & s_{v_{0},a} & s_{a}^2
                         \end{array}\right)\label{eq:time dep matrix}
 \end{align}
 We have placed relevant mathematical details of this derivation in Appendix \ref{ap:emittance math}.
We emphasize that the non-time dependent elements in this matrix are
determined from the initial conditions and therefore
the emittance growth is completely determined by the initial conditions
as one would expect for a deterministic system.  In Fig. 
\ref{fig:emittance growth}, we demonstrate
that this equation perfectly agrees with planar symmetric simulations for various parameters
that satisfy our assumption of laminar conditions. This
perfect agreement is expected as Eq. (\ref{eq:time dependent determinant}) is 
an exact description of emittance growth under laminar flow.  

All initial conditions in Fig. \ref{fig:emittance growth} begin with $0$ emittance
and we set the initial velocity to be linearly related to the initial position, i.e. 
$v_{0,i} = {\bar{v}}_{0,i} +  C (x_{0,i} - {\bar{x}}_{0,i})$.  We call the constant $C$ the chirp,
and if it is unknown it can be obtained from
\begin{align}
  C &= \frac{s_{x,v_x}}{s_x^2}
\end{align}  
The
emittance growth seen in Fig. 1a is then a result of the distribution having a non-linear relation between
the expected momentum at a specified position
and $x$.  Specifically in
non-uniform cases, the $x$-$p_x$ relation becomes non-linear, and our linear assumption
for this relation results in $\eta_x$ being larger than the deviation from the true non-linear $x-p_x$  behavior.   
This in turn results in a growth in the emittance.   Note that we are using emittance here to describe rms emittance, and that
there is a more general understanding of ``true" emittance as it relates to Liouville's theorem in accelerator physics. 
The effects of non-linearity on the rms emittance, again henceforth called emittance, is trivially
calculable from the initial condition through Eq. (\ref{eq:time dependent determinant}), and we
will calculate the expectation of this effect later 
in Sections \ref{sec:exp:gauss} and \ref{sec:exp:bimod} for Gaussian and quadratic bimodal distributions, respectively.
Though we have concentrated on the statistical definition of emittance, 
it is evident that a deeper analysis of deviations from the non-linear $x-p_x$ 
relation would yield more insight into the true momentum spread and
stochastic energy spread.

\begin{figure}
  \centering
   \includegraphics[width=0.3\textwidth]{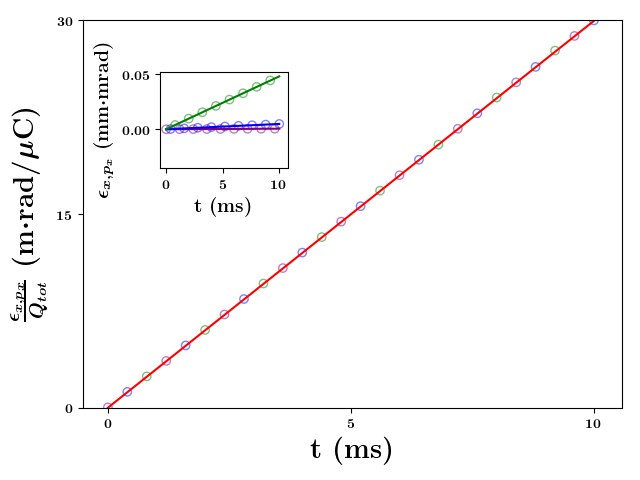}    
  \caption{\label{fig:1D vary a}
  Plots of emittance growth for the similar ensembles of 10k macro-particles under planar symmetry;
  solid lines indicate theory and hollow circles indicate simulation.   
  Ensembles were constructed to have all the same statistics except $s_{a_x}$.
  Specifically, all 
  ensembles were the same 10k macro-particles sampled from a spatially Gaussian distribution with
  standard deviation of $1$ mm and started from rest.  The ensembles differed only 
  by the assumed charges of the macro-particles.
  This difference corresponds to the distribution with different charge 
  densities which in turn results in different $s_{a_x}$ 
  statistics for the ensembles.  
  Notice that when the charge is small (corresponding to low density), 
  emittance for the simulation does not 
  appreciably increase from its initial value of $0$; however, for large charges (corresponding to high density), 
  emittance increases drastically and quickly.}
\end{figure}

The effect of the strength of the self-field within the distribution can also be examined analytically.
This can be done by adjusting the charges on the particles in the simulation for the same initial conditions.
Specifically, we sampled $10k$ particles from an initially Gaussian distribution.  Using the same sample of
initial particle placement but varying the assumed charge per each particle allows us to isolate
the effect of the self-field.  As can be seen in Fig. \ref{fig:1D vary a}, higher
charge densities result in a more rapid emittance evolution.  For the special case examined where the initial
emittance is $0$ and there is no initial velocity, the time constant for this emittance evolution can be shown to be
proportional to the square root of the density, i.e. the timescale is essentially the plasma period.  Likewise,
the slope of the emittance can be shown to be proportional to the square root of the density.  Putting these two 
terms together, for the same period of time the emittance growth will be proportional to the total charge of the distribution
as seen in Fig. \ref{fig:1D vary a}; 
however, ensembles with more complex initial conditions have an emittance whose dynamics have a more 
complex dependence on the density. 

\begin{figure*}
  \centering
  \begin{tabular}{ccc}
    \subfloat[$\sigma_\eta = 0.1 m/s$ ]{\includegraphics[width=0.3\textwidth]{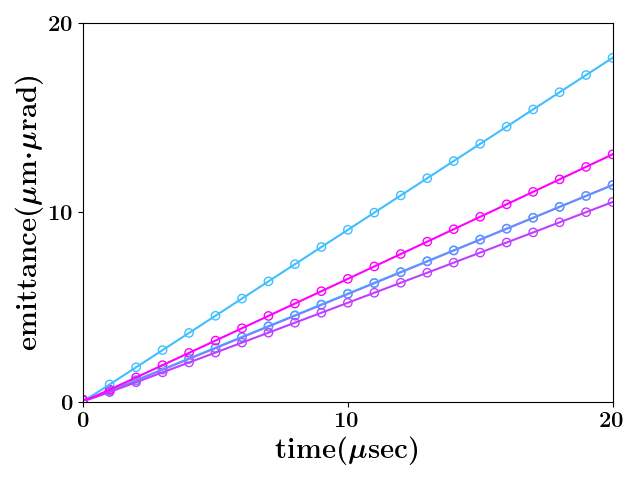}}&
    \subfloat[$\sigma_\eta = 1 m/s$ ]{\includegraphics[width=0.3\textwidth]{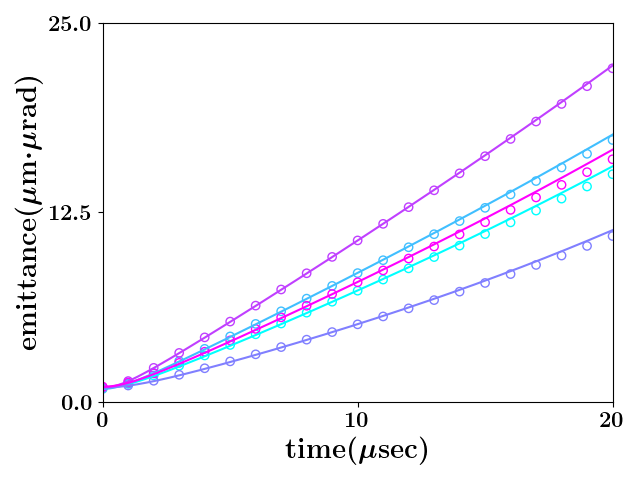}}&
    \subfloat[$\sigma_\eta = 10 m/s$ ]{\includegraphics[width=0.3\textwidth]{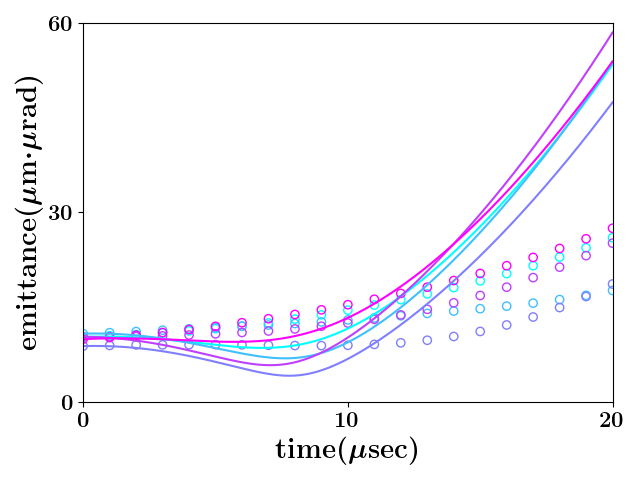}}
  \end{tabular}
  \caption{\label{fig:emittance growth stoch}.  The effect of non-laminar flow on the laminar
  theory of emittance for the evolution of the uniform distribution of $100$ planar particles with
  $\Sigma_{tot}=1.6 \times 10^{-15} ~ \frac{C}{m^2}$ and the initial width of $1 mm$.  
  The parameter $\sigma_\delta$ indicates the standard deviation of the 
  stochastic variable $\delta_i$ with mean $0$ in the equation $v_{0,i} = {\bar{v}}_0 +  C (x_{0,i} - {\bar{x}}_0) + \delta_i$. 
   For (a) $\sigma_\eta = 0.1 \frac{m}{s}$
  the order of particles remains the same in 4 of the 5 simulations, and all simulations are in agreement with
  the theory.  For (b) $\sigma_\delta = 1 \frac{m}{s}$ particle crossover events that change the order of the particles are
  seen in all 5 simulations.  While the theory is still fairly accurate over the time of the simulation, deviation can be seen
  later in the simulation.  For (c) $\sigma_\delta = 10 \frac{m}{s}$ crossover events are again
  seen in all 5 simulations; however, the laminar prediction quickly diverges from the simulated results.}
\end{figure*}

Of course, we could also consider situations where the initial emittance is non-zero due to say a stochastic factor
being included in the initial velocity, i.e. $v_{0,i} = {\bar{v}}_0 +  C (x_{0,i} - {\bar{x}}_0) + \delta_i$
where $\delta_i$ here is a stochastic variable with mean $0$ and units of velocity.  For the simulations presented here, we chose $\delta_i$ from
a Gaussian distribution with standard deviation of $\sigma_{\delta}$.    As can be seen in Fig. \ref{fig:emittance growth stoch}, for small enough
$\sigma_\delta$, the laminar theory correctly predicts the emittance growth, at least for the time period and 
parameters examined.  
However, as expected, the laminar theory begins to diverge from the simulations for moderate $\sigma_\delta$'s,
and for large $\sigma_\delta$, this divergence is almost immediate.  We defer analysis of such ``particle cross-over" effects
tor future studies. 

\section{Statistical dynamics from initial populations}
\subsection{Expected 1D emittance dynamics of a cold bunch}\label{sec:exp:gen}
In the previous section, we obtained an exact expression for the emittance growth that is 
determined entirely from the initial conditions from the sample perspective.  Here we consider the expected emittance growth based on an initial population distribution which is relevant to experiments where the initial population is known, for example at photo-cathodes.  
For the sake of simplicity, we assume that all particles start from rest.  Thus using 
Eq. (\ref{eq:time dependent determinant}) we may write
\begin{align}
  \varepsilon_{x,p}^2 &= \frac{s_{x_0}^2 s_{a}^2 - s_{x_0,a}^2}{c^2} t^2 \label{eq:non rel emsq}
\end{align}
and the expectation of this is
\begin{align}
  \langle \varepsilon_{x,p}^2\rangle &= \frac{\langle s_{x_0}^2 s_{a}^2 \rangle - \langle s_{x_0,a}^2 \rangle}{c^2} t^2\nonumber\\
                                                        &=  \frac{1}{c^2}(\sigma_{x_0}^2\sigma_a^2 - \sigma_{x_0,a}^2 + cov(s_{x_0}^2,s_a^2) - var(s_{x_0,a}))t^2\label{eq:non rel emsq:exp}
\end{align}
analogous to Eq. (\ref{eq:em relation}).

In this section, we assume that the $N$ particles are independently drawn
from a $2$-dimensional phase density where $v_0$ is 0 everywhere---
we refer to this initial rest state as the initial distribution being cold and it is described bythe form,
\begin{align}
  f_{x,p_x,0}(x_0,p_0) &= \rho_0(x_0) \delta(p_0)\label{eq:cold phase}
\end{align}
where $\delta$ is the Dirac delta function and $\rho_0(x_0)$ is the initial spatial density.
Notice that the right hand side of Eq. (\ref{eq:non rel emsq:exp}) is determined solely by the
initial conditions, so the expected emittance in the non-relativistic 1D
model starting from rest can be written as
\begin{align}
  \sqrt{\langle \varepsilon_{x,p}^2\rangle} &= \frac{\sqrt{m_\epsilon^2 + m_{cov}^2}}{c} t\label{eq:em growth gen}
\end{align}
where 
\begin{align}
  m_\epsilon^2 = \sigma_{x_0}^2\sigma_a^2 - \sigma_{x_0,a}^2\label{eq:m_em}
\end{align} 
and 
\begin{align}
  m_{cov}^2 = cov(s_{x_0}^2,s_a^2) - var(s_{x_0,a)}\label{eq:m_stoch}
\end{align}
While $m_\epsilon^2$ is strictly non-negative, $m_{cov}^2$ is not restricted and can have
any value in the real numbers.  Further notice that $m_\epsilon^2$ is determined entirely from the
population perspective whereas  $m_{cov}^2$ represents statistical fluctuations 
amongst of the sample perspective covariance parameters.  In other words, $m_{cov}^2$
can be thought of as a stochastic contribution to the emittance growth due to finite particle number effects.

In the following cases, we choose our distributions such that
either $m_\epsilon = 0$ or $m_\epsilon >> m_{cov}$.  When $m_\epsilon = 0$, 
Eq. (\ref{eq:em growth gen}) reduces to
\begin{align}
  \sqrt{\langle \varepsilon_{x,p}^2\rangle} &= \frac{\lvert m_{cov} \rvert }{c} t \label{eq:lin em2}.
\end{align}
When $m_\epsilon >> m_{cov}^2$, Eq. (\ref{eq:em growth gen}) can be approximated by
\begin{align}
  \sqrt{\langle \varepsilon_{x,p}^2\rangle} &\approx \left(\frac{m_\epsilon}{c}  + \frac{m_{cov}^2}{2cm_\epsilon}\right)t \label{eq:lin em}
\end{align}
As $m_{cov}^2$ can be thought of as a measure of the stochastic contributions to the emittance growth,
define 
\begin{align}
  m_{stoch} &= \begin{cases}
                           \frac{\lvert m_{cov}\rvert}{c},& m_\epsilon = 0\\
                           \frac{m_{cov}^2}{2 cm_\epsilon},& m_\epsilon >> m_{cov}^2
                         \end{cases}
\end{align}
so that $  \sqrt{\langle \varepsilon_{x,p}^2\rangle} \approx \left(\frac{m_\epsilon}{c}  + m_{stoch}\right)t$
in general.

We now use the initial population distribution to obtain $m_\epsilon$.
To obtain this quantity, five expectations need to be calculated
from the population distribution: $\langle x_0\rangle$,
$\langle a \rangle$, $\langle x_0^2 \rangle$, $\langle a^2 \rangle$, and 
$\langle x_0 a \rangle$.     
In the remainder of this section, we show how to obtain these expectations from an
arbitrary population distribution of the form expressed in Eq. (\ref{eq:cold phase}).  
In the following sections, we specifically calculate these expectations, and therefore the associated 
expected emittance,
for the uniform, Gaussian, and the quadratic bimodal distributions,
and we compare these expectations to simulations.

Consider the $3D$ charge density $\rho_q(x,y,z;t)$.  Assuming 
planar symmetry,
this distribution may be decomposed\cite{Zerbe:2018_coulomb_dynamics,Zerbe:2019_relativistic_dynamics,Zerbe:2019_laminar_flow}
\begin{align}
  \rho_q(x,y,z;t) &= \Sigma_{tot} \rho(x;t)
\end{align}
where $\Sigma_{tot}$ is a constant with units of charge per unit area
and $\rho(x;t)$ has units of inverse length and is a probability-like distribution
that normalizes to $1$.  From this distribution, we can calculate the
acceleration of a Lagrangian particle with charge $q$ and mass $m$ at $x_0$:
\begin{align}
  a =  a(x_0) &= \frac{q\Sigma_{tot}}{2m\epsilon_0}\left(\int_{-\infty}^{x_0} \rho_0 d\tilde{x}- \int_{x_0}^{\infty} \rho_0 d\tilde{x}\right)\label{eq:a_s gen}
\end{align}
For a distribution symmetric about $x_0 = 0$, Eq. (\ref{eq:a_s gen}) reduces to
\begin{align}
  a  &= \frac{q\Sigma_{tot}}{m\epsilon_0}\int_{0}^{x_0} \rho_0 d\tilde{x}\label{eq:a_s}
\end{align}
Notice that regardless of the specifics of the 1D real-space distribution,
\begin{align}
  \langle a \rangle = 0\label{eq:exp a}
\end{align}
as required by Newton's third law. 
Furthermore, notice that
the particles are uniformly distributed in acceleration space,
\begin{align}
  \langle a^2 \rangle = \frac{q^2\Sigma_{tot}^2}{12 m^2 \epsilon_0^2} \label{eq:exp asq}
\end{align}
again regardless of the 
specifics of the 1D real-space distribution ---
if this somewhat surprising observation is concerning to the reader,
we encourage the reader to calculate $\langle a^2 \rangle$ themselves
in the cases we discuss shortly.
Finally, we treat the population as a continuum and therefore
\begin{align}
  \langle x_0\rangle = 0
\end{align}  
for distributions symmetric about $x_0 = 0$.
Thus without reference to the specifics of the population distribution profile, we
already know 3 of the 5 required expectations.
The remaining 2 expectations,
$\langle x_0^2\rangle$  and $\langle x_0 a \rangle$, are distribution specific.

\subsection{Uniform}\label{sec:exp:unif}
For the uniform distribution
\begin{align}
  \rho_0(x_0) = \begin{cases}
                           \frac{1}{L}, & -\frac{L}{2} \le z_0 \le \frac{L}{2}\\
                           0, & else
                         \end{cases}
\end{align}
Using this in Eqs. (\ref{eq:a_s}) and (\ref{eq:exp}) we get 
\begin{subequations}
\begin{align}
  a(x_0) &= \frac{q\Sigma_{tot}}{m \epsilon_0}\frac{x_0}{L}\label{eq:unif a}\\
  \langle x_0^2\rangle &= \frac{L^2}{12} \label{eq:unif x0sq}\\
  \langle x_0 a\rangle &= \frac{q\Sigma_{tot} L}{12 m \epsilon_0}\label{eq:unif x0a}
\end{align}
resulting in Eq. (\ref{eq:m_em}) becoming
\begin{align}
  m_\epsilon = 0\label{eq:unif em}
\end{align}
\end{subequations}
Thus, the expected emittance growth from the population theory of the evolution of the uniform distribution is zero
as is generally recognized by the community.  Thus, the emittance growth will be determined by $m_{stoch}$ solely.

\begin{figure}  
  \includegraphics[width=0.4\textwidth]{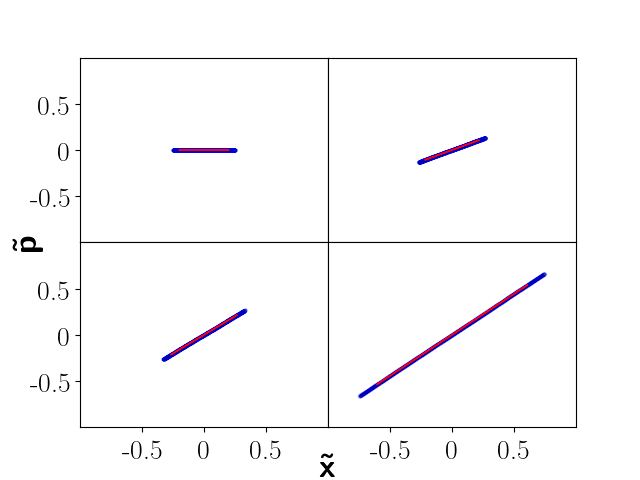}
  \caption{\label{fig:uniform phase-space}
  The normalized ($\tilde{x} = \frac{x}{n_x}$ with $n_x = 0.2~\mu$m, $\tilde{p} = \frac{p}{n_p}$ with 
  $n_p = 2 \times 10^{-7} m c$)
  phase-space of a uniform $M$-shell simulation with $L = 0.1~\mu$m,
  $\Sigma_{tot} = 8 \times 10^{-13} \frac{C}{m^2}$, and $M = 10^4$ at 4 distinct times: 
  (top-left) initial, (top-right) 1 ns, (bottom left) 2 ns, (bottom right) 5 ns.  Blue circles are the
  phase-space positions of the $10^4$ macro-particles, and the red ellipse, which looks like a line
  segment directly over the middle portion of the blue circles, 
  corresponds to 
  the rms ellipse associated with the covariance statistics scaled by $2 \sqrt{2}$. 
  }
\end{figure} 

\begin{figure}  
  \includegraphics[width=0.35\textwidth]{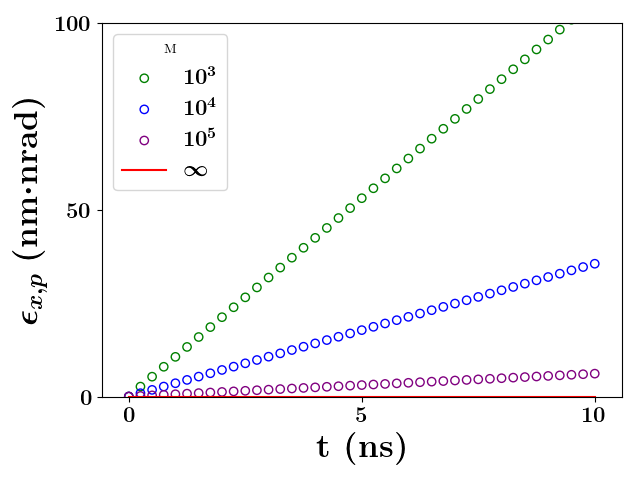}
  \caption{\label{fig:discrete:vs t} The emittance growth of an 
  initially uniform distribution starting from rest 
  simulated with $M$-shell simulations for various
  values of $M$ (circles) or modeled with mean-field theory (red horizontal line
  at y=0).  While
  theory predicts no emittance growth, stochastics of the $M$-shell simulations
  result in apparently linear growth with small slopes that depend on $M$.   
  }
\end{figure}

\begin{figure}  
  \includegraphics[width=0.4\textwidth]{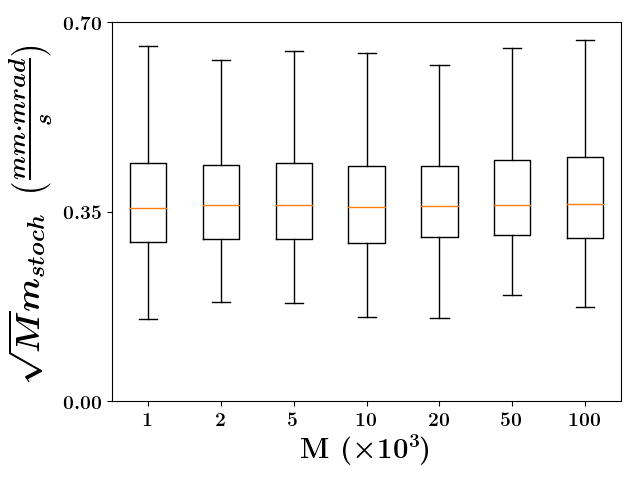}
  \caption{\label{fig:discrete:vs N} Box-plots of the scaled  stochastic slope
  for different values of $M$ generated from 3500 uniform
  $M$-shell simulations.  ANOVA concludes that all simulations have the same scaled
  slope of $0.38 \pm 0.11 \frac{mm \cdot mrad}{s}$.  
  }
\end{figure}

We employed $M$-shell simulation
with $M= 10^4$, $L=0.1 \mu$m, and $\Sigma_{tot} = 8 \times 10^{-13} \frac{C}{m^2}$
 to model the evolution of the uniform distribution.  The parameter $M$ is used instead of $N$
 to emphasize a point --- changes in $M$ are decoupled from changes in the total charge.  Often for $N$-particle
 simulations, each particle across simulations have the same charge resulting in 
 an increase in $N$ indicating an increase in the total charge.  Here assign the charge per particle as 
 $\frac{\Sigma_{tot}}{M}$ so that we can independently vary the number
 of particles and the total charge to more cleanly examine finite particle effects. 
Fig. \ref{fig:uniform phase-space} shows the phase-space at four 
distinct times of one such simulation.  While the phase-space looks
like a straight line, very small variations in the position lead to 
non-zero emittance; in contrast, the population expectations predict exactly
an emittance of zero.  This suggests that this emittance growth is entirely stochastic as anticipated by 
Eq. (\ref{eq:lin em2}).  
This non-zero emittance can be seen in
Fig.  \ref{fig:discrete:vs t} for 3 simulations with different choices of macro-particles but the same total amount of charge.  
The slope, $m_{stoch}$, of this non-mean-field theory emittance growth appears to be linear, as expected,
and is dependent on the choice of $M$.  For clarity, call the quantity $\sqrt{M} m_{stoch}$ the scaled 
stochastic slope. The scaled stochastic slope plotted against $M$ across 500 $M$-shell simulations 
of the initially uniform distribution with these parameters but for each of 7 choices of $M$ ranging from $1k$ to $100k$ can be seen in Fig. \ref{fig:discrete:vs N}. One-way ANOVA \cite{Lane:2017_intro_stats} with $df_1 = 499$ and $df_2 = 3493$ was employed to test the  null hypothesis that all scaled slopes were the same, and the associated F-statistic was 0.78, which accepts the null hypothesis when $\alpha \le 0.05$.  This implies that the slope for rms emittance growth in the uniform planar symmetric distribution is $m_{stoch} = \frac{0.38 \pm 0.11}{\sqrt{M}} \frac{mm \cdot mrad}{s}$.  This result means that
the stochastic and finite-particle effects introduce positive, non-zero emittance growth
for systems characterized by a uniform distribution evolution. 

\subsection{Gaussian}\label{sec:exp:gauss}
We now examine  the Gaussian 1D distribution with standard deviation $\sigma_{x_0}$
\begin{align}
  \rho_0(x_0) = \frac{1}{\sqrt{2 \pi} \sigma_{x_0}} e^{-\frac{x_0^2}{2 \sigma_{x_0}^2}}
\end{align}
Using this in Eqs. (\ref{eq:a_s}) and (\ref{eq:exp}) we get 
\begin{subequations}
\begin{align}
  a(x_0) &= \frac{q\Sigma_{tot}}{2 m \epsilon_0}\text{erf}\left(\frac{x_0}{\sqrt{2}\sigma_{x_0}}\right)\\
  <x_0^2> &= \sigma_{x_0}^2\\
  <x_0 a> &= \frac{q\Sigma_{tot} \sigma_{x_0}}{2 m \sqrt{\pi} \epsilon_0}
\end{align}
resulting in Eq. (\ref{eq:m_em}) becoming
\begin{align}
  m_\epsilon &= \sqrt{\frac{1}{12} - \frac{1}{4 \pi}}\frac{q \Sigma_{tot} \sigma_{x_0}}{m \epsilon_0} \nonumber\\
    &\approx 0.0613  \frac{q \Sigma_{tot} \sigma_{x_0} }{m \epsilon_0} \label{eq:G non emitttance}
\end{align}
\end{subequations}

\begin{figure}  
  \includegraphics[width=0.4\textwidth]{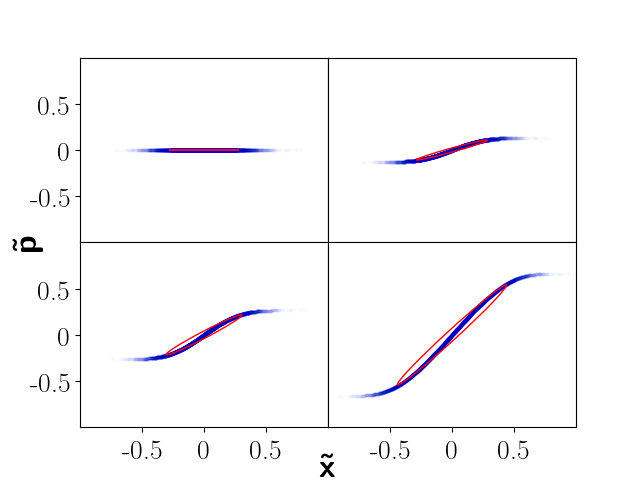}
  \caption{\label{fig:gaussian phase-space}
  The normalized ($\tilde{x} = \frac{x}{n_x}$ with $n_x = 0.5~\mu$m, $\tilde{p} = \frac{p}{n_p}$ with
  $n_p = 2 \times 10^{-7} m c$)
  phase-space of a Gaussian $M$-shell simulation with $\sigma_{x_0} = 0.1~\mu$m,
  $\Sigma_{tot} = 8 \times 10^{-13} \frac{C}{m^2}$, and $M = 10^4$ at 4 distinct times: 
  (top-left) initial, (top-right) 1 ns, (bottom left) 2 ns, (bottom right) 5 ns.  Blue circles are the
  phase-space positions of the $10^4$ macro-particles, and the red ellipse 
  corresponds to the
  the rms ellipse associated with the covariance statistics scaled by $2 \sqrt{2}$. 
  }
\end{figure} 

\begin{figure}  
  \includegraphics[width=0.35\textwidth]{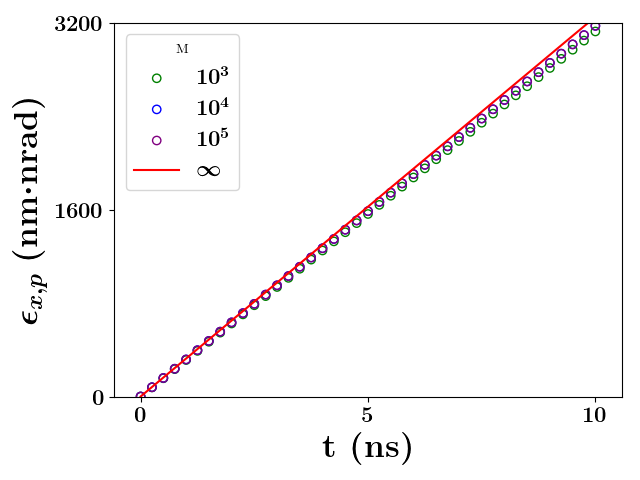}
  \caption{\label{fig:gaussian discrete:vs t} The emittance growth of an 
  initially Gaussian distribution starting from rest 
  simulated with $M$-shell simulations for various
  values of $M$ (circles) or modeled with mean-field theory (red line).  
  All simulations are fairly similar to theory suggesting the non-linearity 
  captured by the mean-field model dominates
  stochastic effects for the Gaussian distribution.
  }
\end{figure}

We again employed $M$-shell simulation
with $M = 10^4$, $\sigma_{x_0} =0.1 \mu$m, and $\Sigma_{tot} = 8 \times 10^{-13} \frac{C}{m^2}$
 to model the evolution of Gaussian distribution.
Unlike the uniform distribution, Fig. \ref{fig:gaussian phase-space} shows that 
visible non-linearity arises during the simulation.  Simulations with
3 different $M$ with the same total charge agree well with theory as can be seen in Fig. 
\ref{fig:gaussian discrete:vs t} allowing us to make the assumption that $m_\epsilon^2 >> m_{cov}^2$. 

\begin{figure}  
  \includegraphics[width=0.4\textwidth]{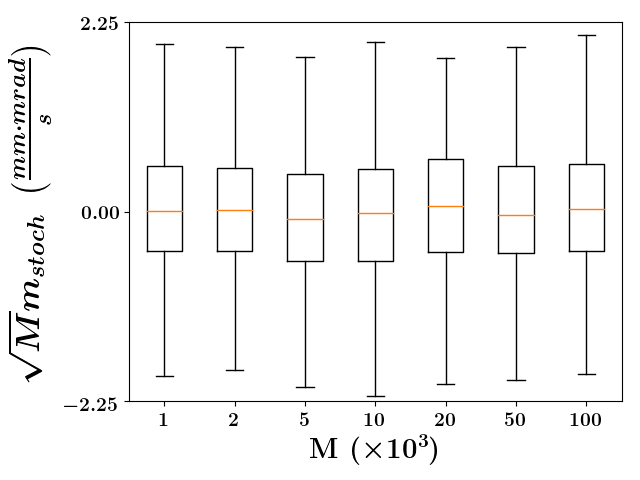}
  \caption{\label{fig:gaussian discrete:vs N} Box-plots of the scaled stochastic slope
  for different values of $M$ generated from 3500 Gaussian
  $M$-shell simulations.  ANOVA concludes that all simulations have the same scaled remainder
  slope of $0 \pm 0.8 \frac{mm \cdot mrad}{s}$.  
  }
\end{figure}

As expected, small variations of the slope are also evident in Fig. 
\ref{fig:gaussian discrete:vs t}.  To explore this variation, we subtracted
the $\frac{m_\epsilon}{c} t$ from the emittance leaving us with $m_{stoch}$.
Again, we examine the scaled stochastic slope, $\sqrt{M} m_{stoch}$.  The scaled 
stochastic slope plotted against $M$ across 500 $M$-shell, same total charge simulations 
of the initially Gaussian distribution with these parameters but
for each of 7 choices of $M$ ranging from $1k$ to $100k$ can be seen in Fig. 
\ref{fig:gaussian discrete:vs N}.
One-way ANOVA with $df_1 = 499$ and $df_2 = 3493$ was employed to test the
null hypothesis that all scaled slopes were the same, and the associated
F-statistic was 1.95, which still accepts the null hypothesis when $\alpha \le 0.05$.  
This implies that 
the stochastic contribution to the slope for emittance growth 
in the Gaussian planar symmetric distribution is 
$m_{stoch} = \frac{0 \pm 0.8}{\sqrt{M}} \frac{mm \cdot mrad}{s}$.  For comparison,
the mean-field slope is $0.33 \frac{mm \cdot mrad}{s}$ with these parameters which does have the property 
$m_\epsilon >> m_{stoch}$ for $M \ge 1000$ as expected; that is the non-linear mean-field effects dominate emittance growth of systems characterized by the Gaussian distribution for sufficient $M > 1000$.

\subsection{Quadratic Bimodal}\label{sec:exp:bimod}

For the quadratic bimodal distribution
\begin{align}
  \rho_0(x_0) = \begin{cases}
                           \frac{12}{L^3} x_0^2, & -\frac{L}{2} \le x_0 \le \frac{L}{2}\\
                           0, & else
                         \end{cases}
\end{align}
Using this in Eqs. (\ref{eq:a_s}) and (\ref{eq:exp}) we get 
\begin{subequations}
\begin{align}
  a_s(x_0) &= \frac{4 q\Sigma_{tot}}{m \epsilon_0}\frac{x_0^3}{L^3}\\
  \langle x_0^2 \rangle &= \frac{3 L^2}{20}\\
  \langle x_0 a \rangle &= \frac{3 q\Sigma_{tot} L}{28 m \epsilon_0}
\end{align}
resulting in Eq. (\ref{eq:m_em}) becoming
\begin{align}
  m_\epsilon &=  \sqrt{\frac{1}{12} - \frac{15}{196}}\frac{q \Sigma_{tot} \sigma_{x_0} }{m \epsilon_0} \nonumber\\
    &\approx 0.0825  \frac{q \Sigma_{tot} \sigma_{x_0} }{m \epsilon_0} \label{eq:bi non emitttance} 
\end{align}
\end{subequations}

\begin{figure}  
  \includegraphics[width=0.4\textwidth]{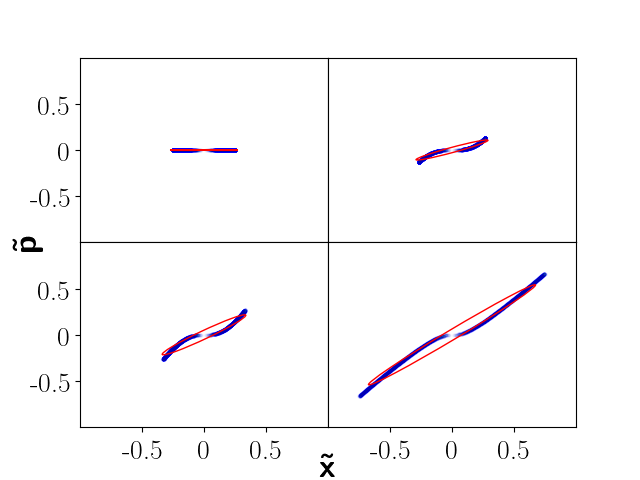}
  \caption{\label{fig:bimodal phase-space}
  The normalized ($\tilde{x} = \frac{x}{n_x}$ with $n_x = 0.2~\mu$m, $\tilde{p} = \frac{p}{n_p}$ with
  $n_p = 2 \times 10^{-7} m c$)
  phase-space of a bimodal $M$-shell simulation with $L = 0.1~\mu$m,
  $\Sigma_{tot} = 8 \times 10^{-13} \frac{C}{m^2}$, and $M = 10^4$ at 4 distinct times: 
  (top-left) initial, (top-right) 1 ns, (bottom left) 2 ns, (bottom right) 5 ns.    
  Blue circles are the
  phase-space positions of the $10^4$ macro-particles, and the red ellipse corresponds to the
  the rms ellipse associated with the covariance statistics scaled by $2 \sqrt{2}$.
  }
\end{figure}

\begin{figure}  
  \includegraphics[width=0.35\textwidth]{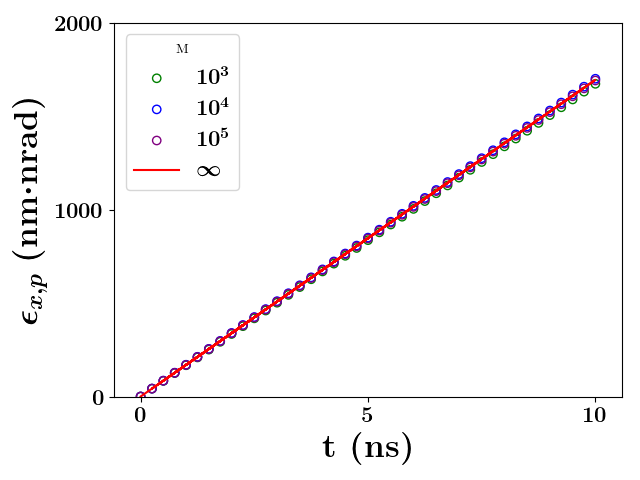}
  \caption{\label{fig:bimodal discrete:vs t} The emittance growth of an 
  initially bimodal distribution starting from rest 
  simulated with $M$-shell simulations for various
  values of $M$ (circles) or modeled with mean-field theory (red line).
  All simulations are fairly similar to theory suggesting the non-linearity 
  captured by the mean-field model dominates
  stochastic effects for the bimodal distribution.     
  }
\end{figure}

We employed $M$-shell simulation
with $M = 10^4$, $L=0.1 \mu$m, and $\Sigma_{tot} = 8 \times 10^{-13} \frac{C}{m^2}$
 to model the evolution of the bimodal distribution.
Fig. \ref{fig:bimodal phase-space} shows the phase-space at four 
distinct times of one such simulation, and this phase distribution has
a visible non-linear kink about the origin, although in the opposite direction to the kink seen in the Gaussian distribution, that is related to the non-zero emittance growth of systems decribed by this distribution.
Simulations with 3 different $M$ with the same total charge agree well with theory as can be seen in Fig. 
\ref{fig:bimodal discrete:vs t} again suggesting that we make the assumption $m_\epsilon^2 >> m_{cov}^2$.

\begin{figure}  
  \includegraphics[width=0.4\textwidth]{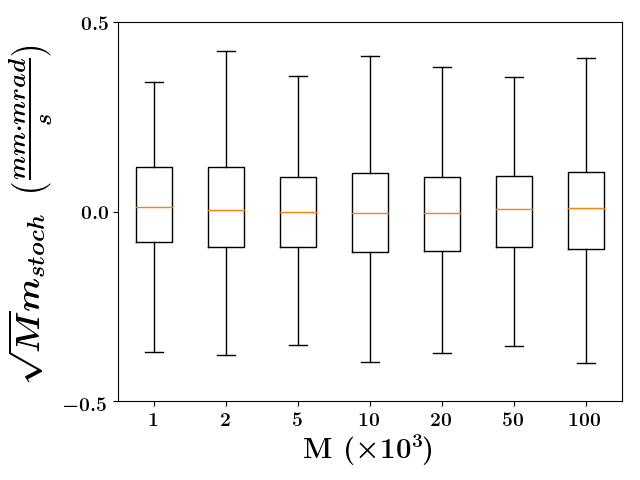}
  \caption{\label{fig:bimodal discrete:vs N} Box-plots of the scaled stochasitc slope
  for different values of $M$ generated from 3500 uniform
  $M$-shell simulations.  ANOVA concludes that all simulations have the same scaled
  slope of $0.38 \pm 0.11 \frac{mm \cdot mrad}{s}$.  
  }
\end{figure}

We again subtracted $\frac{m_\epsilon}{c} t$ from the emittance leaving us with $m_{stoch}$,
and we examined the scaled stochastic slope, $\sqrt{M} m_{stoch}$.  The scaled 
stochastic slope plotted against $M$ across 500 $N$-shell, same total charge simulations 
of the initially bimodal distribution with these parameters but
for each of 7 choices of $M$ ranging from $1k$ to $100k$ can be seen in Fig. 
\ref{fig:bimodal discrete:vs N}.
One-way ANOVA with $df_1 = 499$ and $df_2 = 3493$ was employed to test the
null hypothesis that all scaled slopes were the same, and the associated
F-statistic was 0.83, which accepts the null hypothesis when $\alpha \le 0.05$.  
This implies that 
the stochastic contribution to the slope for emittance growth 
in the bimodal planar symmetric distribution is 
$m_{stoch} = \frac{0.01 \pm 0.15}{\sqrt{M}} \frac{mm \cdot mrad}{s}$, which is again
much smaller than $m_\epsilon = 0.17  \frac{mm \cdot mrad}{s}$
for sufficient $M$.  
Notice that the distribution of stochastic slopes is again roughly symmetric about zero similar to the Gaussian case
but with a smaller standard deviation.

\section{The sample perspective and kinetic energy}\label{sec:KE}

Kinetic energy plays a central role in the sample perspective as kinetic energy can be decomposed using 
ensemble statistics.  Namely,
\begin{align}
  KE &= \sum_{i=1}^N \sum_{j\in\{x,y,z\}} \frac{p_j^2}{2m}\nonumber\\
        &= \sum_{j\in\{x,y,z\}} KE_j\label{eq:KE}
\end{align}
where $KE_j = \frac{p_j^2}{2m}$ is the portion of the kinetic energy contained in the $j^{th}$ dimension.  
For the planar model, only the kinetic energy of one of these dimensions is relevant.
Specifically, the portion of the kinetic energy along $x$ can be written as
\begin{align}
  KE_x &= \sum_{i=1}^N \frac{p_x^2}{2m} - N \frac{{\bar{p}}_x^2}{2m} + N \frac{{\bar{p}}_x^2}{2m}\nonumber\\
            &= \sum_{i=1}^N\frac{(p_x-{\bar{p}}_x)^2}{2m} + N KE_{CoM,x}\nonumber\\
            &= N \frac{s_{p_x}^2}{2m} + N KE_{CoM,x}\label{eq:KE exact}
\end{align}
where $KE_{CoM,x} = \frac{{\bar{p}}_x^2}{2m}$ is the kinetic energy of a single particle traveling with the mean 
momentum of the distribution, i.e. the center of mass (CoM) motion of the ensemble.  Notice this
equality is exact in the sample perspective, and any discussion of
energy, so central to physics, thus necessitates an understanding of the momentum variance.

This energy can be further decomposed using the standard notation of emittance.
Specifically, using $\mu_x = \frac{s_{x,p_x}}{s_x}$ and $\eta_x^2 = s_{p_x}^2 - \mu_x^2$, the kinetic energy can be
written as
\begin{align}
  KE_x &= N \frac{\eta_x^2}{2m} + N \frac{\mu_x^2}{2m} + N KE_{CoM,x}\label{eq:KE modes}
\end{align}
While this may not look as if the statistical definition of
emittance is used here, remember that emittance is $\varepsilon_{x,p_x} = \frac{1}{mc} s_x \eta_x$.  Therefore,
if we analyze the physical processes affecting the energy term,  $N \frac{\eta_x^2}{2m}$, we can develop a deeper understanding of emittance dynamics. 
Specifically, $N \frac{\mu_x^2}{2m}$ can be thought of as a mode 
that contains the kinetic energy of the linear flow motion of the distribution
and $N \frac{\eta_x^2}{2m}$ can be thought of as the mode that contains the 
kinetic energy remaining when the linear flow energy is removed, i.e. the kinetic energy fluctuations or "heat".

One caveat, though, is that $N \frac{\mu_x^2}{2m}$ is an estimate of the flow energy using the linear $x-p_x$ relation that is assumed in almost all of beam physics, and this approximation is sometimes questionionable.  Fortunately for the planar model
we can exactly calculate the kinetic energy of the distribution.  Quite remarkabley, the kinetic energy
of a planar distribution at time $t$ starting from rest at time $0$ is given by Eq. (\ref{eq:KE exact}):
\begin{align}
  KE &= \frac{N m}{2} s_a^2 t^2\label{eq:ke stat}
\end{align}
where $s_a^2 =  \frac{q^2\Sigma_{tot}^2}{12 m^2 \epsilon_0^2}$ by Eq. (\ref{eq:exp a})
and Eq. (\ref{eq:exp asq}).  This result is exact regardless of the initial density profile of the distribution. The universality of this result is pecular to the 1-D constant acceleration case which enable analytic insight that is hard to extract in more complex systems.   A calculation of the linear flow energy yields,
\begin{align}
  N \frac{\mu_x^2}{2m} &= \frac{Nm}{2} \frac{s_{x_0,a}^2 + s_{x_0,a}s_a^2 t^2 + \frac{1}{4}s_{a}^4 t^4}{s_{x_0}^2 + s_{x_0,a} t^2 + \frac{1}{4} s_a^2 t^4} t^2\label{eq:exp energy:gen}
\end{align}
Notice that this estimate is profile specific.  Namely, $s_{x_0,a}$ and $s_{x_0}^2$ differ depending on the profile.
We have already calculated the expectation of these
covariances under the uniform, Gaussian and bimodal distributions in Sections \ref{sec:exp:unif}, 
\ref{sec:exp:gauss}, and \ref{sec:exp:bimod}, respectively, so no
additional calculation is necessary.  For the uniform case, notice we have $s_{x_0,a} = s_{x_0} s_{a}$.  Putting
this into Eq. (\ref{eq:exp energy:gen}), the uniform case linear expansion energy reduces to Eq. (\ref{eq:ke stat}); that is,
the linear flow energy exactly captures the kinetic energy evolution of the uniform distribution, so when the distribution expands,  no energy is transferred to the heating mode.  Furthermore, for 
long time, the linear flow energy of the non-uniform distributions asymptote to  Eq. (\ref{eq:ke stat}); however,
for the non-uniform distributions at any real time, $N \frac{\mu_x^2}{2m} < KE$.  That is, even though these distributions have
the same energy in the expansion mode as the uniform distribution, the statistics we use underestimates the
amount of kinetic energy associated with expansion.

\begin{figure}  
  \includegraphics[width=0.35\textwidth]{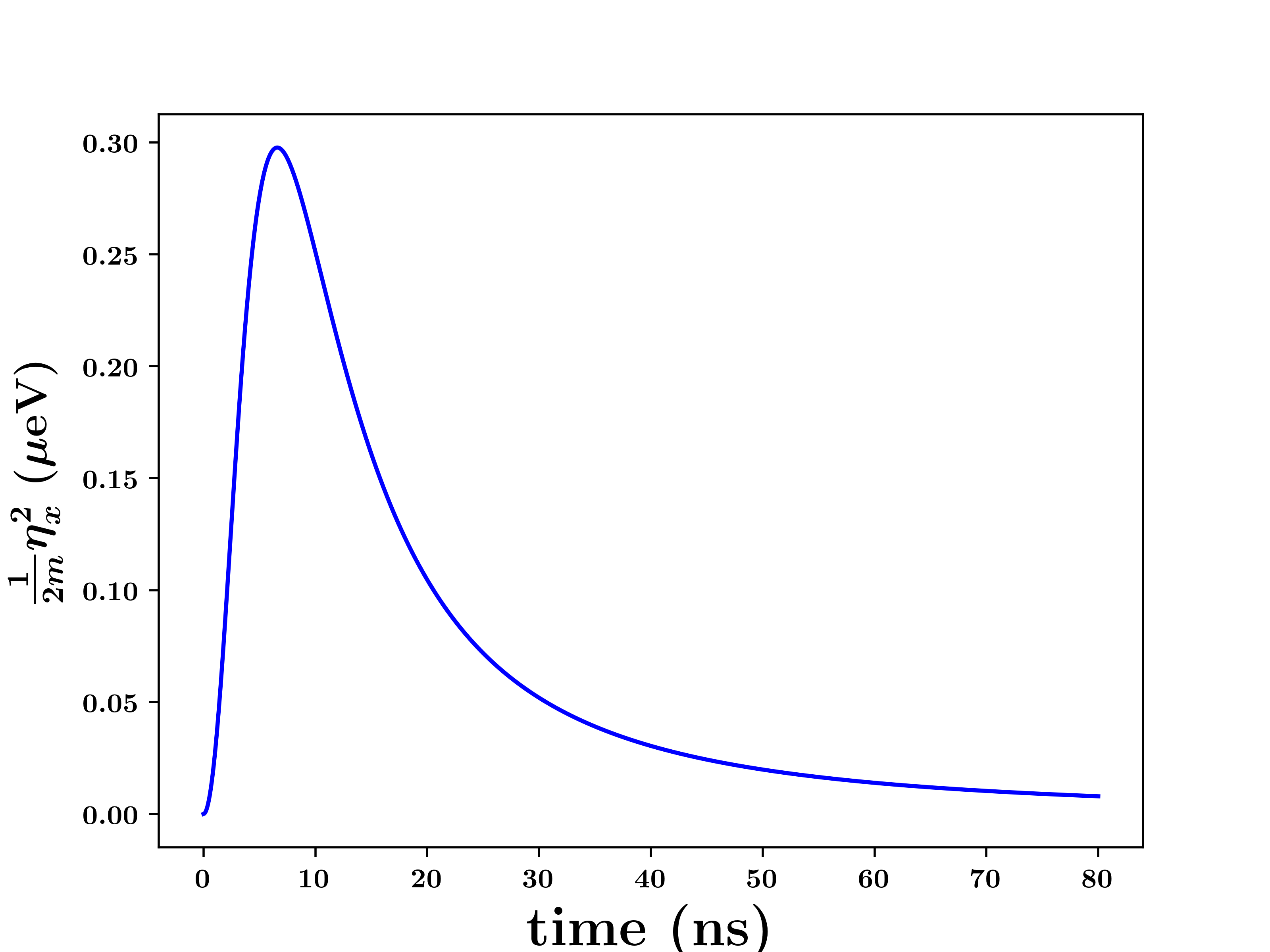}
  \caption{\label{fig:Gaussian heating} The evolution of the linear heat along x, $\frac{1}{2m} \eta_x^2$,
  for an ensemble of 
  $10k$ planar particles initially with Gaussian-spatial distribution
  with density $\Sigma_{tot} = 8 \times 10^{-13} \frac{C}{m^2}$
  and initial width of $0.1 ~ \mu$m.  Notice that kinetic energy is initially
  released as heat into the distribution, and then this heat is rapidly lost to expansion.  
  This is qualitatively similar to the evolution of the Mean Transverse Energy 
  (mathematically identical what we are calling the linear heat along $x$)
  seen by Maxson et. al\cite{Maxson:2013_DIH}.  We note that Maxson et al.'s
  explanation for this effect is a phenomenological $3D$ model that captures the phenomenon
  with fitted parameters while we present an analytic
  $1D$ model that does not yet quantitatively capture the effect described by Maxson et. al.     
  }
\end{figure}

Since the heat is the left over energy, any underestimation of the linear flow energy means that a portion of the heat,
$N \frac{\eta_x^2}{2m} = N \frac{p_x^2 - \mu_x^2}{2m}$,
is actually linear flow energy.  In fact, we can easily obtain a functional form for the 
heat when the planar distribution starts from rest with
\begin{align}
  \frac{N}{2m} \eta_x^2 &= \frac{Nm}{2} \frac{c^2\varepsilon_{x,p_x}^2}{s_{x}^2}\nonumber\\
                                     &= \frac{Nm}{2}\frac{s_{x_0}^2s_a^2 - s_{x_0,a}^2}{s_{x_0}^2 + s_{x_0,a} t^2 + \frac{1}{4} s_a^2 t^4} c^2 t^2\label{eq:KE_fluc ev}
\end{align}
Of course, this is simply the complementary energy to the linear flow energy in Eq. (\ref{eq:exp energy:gen}) as their sum gives 
Eq. (\ref{eq:ke stat}) in all cases.
The evolution of $\frac{N}{2m} \eta_x^2 $ of $10k$ planar particles initially with Gaussian-spatial distribution
with density $\Sigma_{tot} = 8 \times 10^{-13} \frac{C}{m^2}$ can be seen in Fig. \ref{fig:Gaussian heating}.
All initially cold-$1D$ distributions exhibit this characteristic, single-hump evolution.
This release of heat from the $1D$ model is qualitatively the same as seen by Maxson et. al in their 
study of disorder induced heat\cite{Maxson:2013_DIH} in fully $3D$ systems; however, additional concerns
are present in the full $3D$ model, and we will discuss these concerns shortly.

\begin{figure}
\centering
   \includegraphics[width=0.5\textwidth]{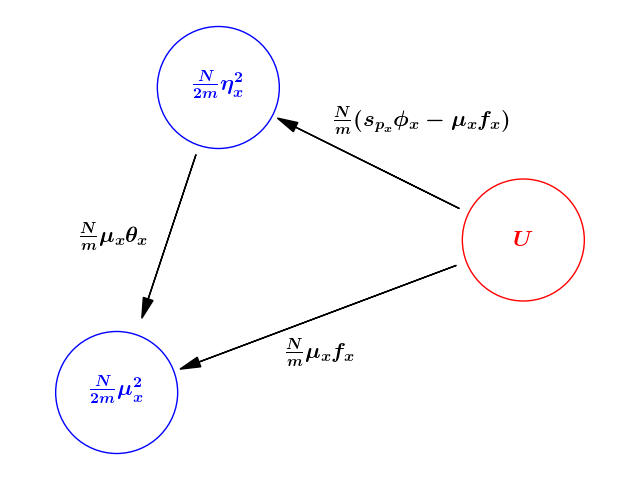}
  \caption{\label{fig:1D mode}
  Exact schematic of the non-relativistic kinetic energy and potential energy of a generic $1D$ system from the sample
  perspective.  The labelled circles represent modes where energy can be found, and the arrows represent power channels across
  which energy can be exchanged between the modes.  Note that the label on the power channel may be negative thus reversing the
  direction of energy flow.  $U$ represents potential energy, $\eta_x = \sqrt{s_{p_x}^2 - \mu_x^2}$,
  $f_x = \frac{s_{x,F_x}}{s_x}$, $\phi_x =  \frac{s_{p_x,F_x}}{s_{p_x}}$, and $\theta_x = \frac{1}{m} \frac{\eta_x^2}{s_x}$.  
  }
\end{figure}

However, before we move onto the discussion of additional $3D$ effects, we first further discuss the
evolution of the three modes we recognize in our model: the potential energy ($U$), the
linear flow mode, and the linear heat mode.  Taking the time derivative of the linear flow energy, we obtain
\begin{align}
  \frac{d}{dt} \frac{N}{2m} \mu_{x}^2 &= \frac{N}{m} \mu_x {\dot{\mu}}_x \nonumber\\
                                        &= \frac{N}{m} \left(\mu_x f_x + \mu_x \theta_x\right)
\end{align}
where $f_x = \frac{s_{x,F_x}}{s_x}$ can be thought of as the self-force at one standard deviation within the spatial-force distribution
as expected from linear regression in the same way $\mu_x = \frac{s_{x,p_x}}{s_x}$ is the corresponding momentum, and
$\theta_x = \frac{1}{m} \frac{\eta_x^2}{s_x}$ can be thought as some kind of effective heating force.
Taking the time derivative of the heat we obtain
\begin{align}
  \frac{d}{dt} \frac{N}{2m} \eta_{x}^2 &= \frac{N}{2m} \frac{d}{dt} \left(s_{p_x,p_x} - \mu_x^2\right) \nonumber\\
                                        &= \frac{N}{m} \left(s_{p_x} \phi_x - \mu_x f_x - \mu_x \theta_x\right)
\end{align} 
 where $\phi_x =  \frac{s_{p_x,F_x}}{s_{p_x}}$ is analogous to $f_x$ and $\mu_x$ except for it is the linear prediction
 of the momentum-force relation.  Notice that these equations are simply two of the three statistical kinematics equations
 used to derive the envelope equations.  Inspection of these terms allows us to isolate the three 
 ``power channels" through which energy
 is transferred between these modes and the potential energy as can be seen in Fig. \ref{fig:1D mode}.

If we consider a non-interacting model, there is no force.  In such a model, $f_x = 0$ and 
$\phi_x =  0$ by definition.  The term $\theta_x$ is only zero if the bunch has no heat (and hence zero emittance), 
otherwise it is not zero.  So for any non-interacting model with non-zero emittance, there is energy
flow between the linear heat and the linear flow energy, given by  $\frac{N}{2m} \mu_x \theta_x$, as the bunch evolves.
This is the only power channel available to the non-interacting model, and we will call this channel the non-interacting
channel.

When we consider interacting models, we note that the power channel labelled by $\frac{N}{m} (s_{p_x} \theta_x - \mu_x f_x)$
can also be written as $\frac{N}{2} mc^2 \frac{1}{s_x^2} \frac{d\varepsilon_{x,p_x}^2}{dt}$ -- that
is, this energy flows across this power channel only if emittance changes.  We will call this channel the 
emittance change channel.  Therefore, the envelope equations,
which conserve emittance, do not have any energy exchange directly between the potential and the heat
through the emittance change channel.  Instead,
they have the non-interacting channel and the additional power channel between the 
potential and the linear flow energy, which we will call the flow channel.  
Of course, if we were to use a model that does not conserve emittance, all three channels; non-interacting,
flow, and emittance change,
would be accessible --- this is the case we see in the planar model.

We can now qualitatively 
describe what is happening in Fig. \ref{fig:Gaussian heating}.  When the planar distribution starts from rest, the potential
is converted to linear flow energy and linear heat through the flow and emittance change channels, respectively.  
Again, the emittance change channel occurs due to deviations from linearity.  However, as the distribution
begins to expand and heat up, $\mu_x$ and the heat both get larger.  Eventually, these two values are sufficient to result
in the non-interacting channel, $\frac{N}{2m} \mu_x \theta_x$, becoming larger than the emittance change channel resulting
in a depletion of the linear heat to linear flow energy despite the fact that the emittance channel remains non-zero.

\begin{figure}
\centering
   \includegraphics[width=0.5\textwidth]{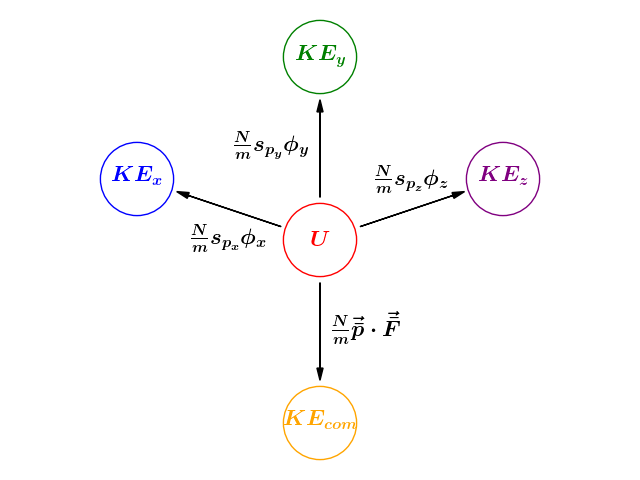}
  \caption{\label{fig:3D mode}
  Exact schematic of the non-relativistic kinetic energy and potential energy of a generic $3D$ system from the sample
  perspective.  The labelled circles represent modes where energy can be found, and the arrows represent power channels across
  which energy can be exchanged between the modes.  Note that the label on the power channel may be negative thus reversing the
  direction of energy flow.  $U$ represents potential energy, $KE_{com} = N \sum_{j\in\{x,y,z\}} \frac{{\bar{p}}_{j}^2}{2m}$  is the kinetic energy
  of the center of mass, and $KE_j = \frac{s_{p_j}^2}{2m}$ for $j\in\{x,y,z\}$.
  Notice that the center of mass mode could be split up into three modes, 
  $x$, $y$, and $z$, but we have not done so here to simplify the schematic as we are not concerned with center of mass motion in this manuscript.  Further notice that 
  all inter-dimensional energy transfer needs to pass through the potential --- i.e. there needs to be an interaction between particles where the energy can be stored.  Finally, $KE_j$ can be split up as in Fig. \ref{fig:1D mode} for each $j$.}
\end{figure}

So this brings us back to why Fig. \ref{fig:Gaussian heating} is so similar to the plot of disorder induced heating
seen at least by Maxson et. al.\cite{Maxson:2013_DIH}.  The reason is because the schematic in Fig. \ref{fig:1D mode}
is still correct in $3D$ with a couple of modifications.  First, instead of just $KE_x$, we now have $KE_y$ and $KE_z$,
which can be split up similarly.
The only way to transfer energy between the dimensions is through the potential.  A full schematic of this picture is seen in
Fig. \ref{fig:3D mode}.  Secondly, as the force in $3D$ is proportional to $\frac{1}{r^2}$, displacements from the reference
distribution, which is no longer uniform except in the continuous case, will affect the potential.  We will examine
this potential difference both from stochastic effects and from the global profile in future work, 
but we point out that the release of this additional potential is analogous to
the non-linear effects already seen in $1D$.  Finally, in the planar model, the kinetic energy of the bunch increases toward
infinity whereas the $3D$ dynamics quickly deplete the potential energy.
So while the $1D$ effects will not capture this additional disorder 
induced heating, they do evolve similarly.

\section{Discussion}\label{sec:discussion}

Here we have presented an approach to non-equilibrium beam dynamics we are calling the sample perspective.
By changing the focus of analysis from the expectation operator dependent on the population density 
to the mean operation on a specific sample, we are able to obtain exact equations for the time evolution of the
sample statistics.  
The theoretical advantage of the sample perspective is that it yields new analytic approaches, useful and complementary physical insight, and it may lead to a systematic calculation of all moments though here we focus on the second moment.  Moreover, for well behaved distributions,  knowledge of all moments provides
an effective analysis of non-equilibrium distributions, provided they are well behaved.
More standard non-equilibrium dynamic approaches, 
like Vlasov's equation or the 
BKGY hierarchy of equations can be computationally difficult and require different and complementary analytic approaches.  As seen here, the sample perspective
supplies surprisingly simple exact descriptions for the case of laminar flows in one dimension, and a number of additional results including results for higher dimensional models are being developed.  

Returning to the difference between the population and sample perspectives, we point out the collisions present in the population perspective are absent from the sample perspective.   Collisions are required to fully capture the interactions in the population theory, but the interactions are exactly captured through the interparticle forces in the sample paradigm.  Nevertheless there are many ways in which the collisions required in the population perspective can be estimate by using the sample calculations, for example calculations of the variance of the emittance in the sample perspective may be used to estimate the collision term in the population perspective, as will be described elsewhere.  It is important to distinguish between the collision term in the population perspective and scattering events which are exactly captured in the sample perspective, provided the exact force is used.   Moreover in a classical calculation, and assuming a system that is not chaotic, the emittance dynamics is deterministic provided the initial conditions are known exactly, as noted by Sacherer\cite{Sacherer:1971_envelope}.

Of course, collision-less is not synonymous with emittance conserving; specifically, we saw emittance growth
in our planar symmetric model.  Easy to appreciate are the global non-linearities that arise in phase space
that are due to non-linearities that arise from real-space planar distribution being non-linear.  Anderson calculated 
the effect of such
non-linearities using the population perspective well over 3 decades ago\cite{Anderson:1987_emittance}, and 
the community is quite aware of their role\cite{Buon:1992_phase_space}.
Of course, we saw, especially in the 
analysis of uniform planar-symmetric distributions, that stochastic effects can also play a significant role; however, stochastic
effects, as presented in the planar symmetric model, are in fact just another form of non-linearity. 
It is common to think of the force in terms of the mean field force, but finite particle effects, even in the planar symmetric model,
result in deviations from the expected variances and covariances.  In turn, these deviations introduce additional non-linear effects.
In the case of the mean-field uniform distribution, the introduced non-linearities are dominant as the mean-field non-linearities are
completely absent.  In the case of the Gaussian and quadratic bimodal distribution examined in this work, the additional non-linearties
are much smaller than the mean-field non-linear contributions.  Furthermore, these stochastic non-linearities may in fact make such
non-uniform distributions more-linear.  So it should come as no surprise that the average uniform emittance slope increase 
due to stochastic effects is non-zero but the analogous term for the non-uniform cases is symmetric about zero.
In fact, these stochastic non-linearities
can be obtained analytically, but such analysis is involved and beyond the scope of this introductory presentation.  We will
present such theory in future work.

Again, the use of the covariance statistics is inherently linear as $\frac{s_{x,p_x}}{s_x^2}$ is the slope of the line of best fit relating $x$ and $p_x$.   So the (statistical) emittance growth can be argued to be an artifact of the linear statistics used to define the second moments.  This is an important feature of the emittance utilized in the field of accelerator physics, and is due to the fact that
linear optics are generally used when manipulating charged particle beams, and therefore, this linear
statistic is appropriate for such standard techniques.  Our second counter argument is theoretical. 
The most natural nonlinear choice for
the function relating the position and momentum at $t$ would be to choose the average momentum at a specific
position.  Doing so eliminates the mean-field portion of the slope,
$m_{mean}$,  from  Eq. (\ref{eq:m_stoch}).  That is, under this alteration of the statistics,
the expected emittance growth would follow $m_{stoch}$, which represents how the sample
distribution differs from the population distribution, solely.  As we can largely separate out $m_{stoch}$ 
from the underlying distribution, such non-linear treatment does does not provide any
additional physical insight.
Of course, we could adopt the strategy to remove the non-linearities in the measure by 
finely-slicing and measuring the emittance of each slice as an alternative method of getting rid of the non-linearities.  
However, again this reduces to calculating $m_{stoch}$.  Therefore, we conclude that the effects on emittance
of these non-linearities should be separated out in future theoretical treatments, and a better understanding 
of $m_{stoch}$ is needed.

We close this section by emphasizing again that the sample perspective is exact if the interparticle force and initial statistical configuration are exactly known;
specifically, the full 3D statistical kinematic equations, a generalization of
Eqs. (\ref{eq:dsx^2dt}) - (\ref{eq:dspx^2dt}), exactly describes the evolution of the statistics 
as alluded to by Sacherer.  Central in such analyses is the force, and the force is where the sample and population perspectives
are most divergent.  Specifically, in the sample perspective  the force on each particle  is exactly specified 
by the $N$ values of the particles in phase space; in the population perspective, the force
is ill-defined although often approximated by the mean field force\cite{Sacherer:1971_envelope}.     
The role of the force can be easily seen in the two cases we examined here. 
Namely, the force is trivially calculated from the ensemble as it is either exactly 0 (in the non-interacting case) or 
is trivially constant for each particle (the laminar planar symmetric case).  This exact specification of the force is why 
these situations lead to exact solutions to the evolutions of the statistics.  On the other hand in  full $3D$ dynamics, 
the force is much more complicated and instead needs to be approximated, and one such 
approximation, the mean-field force approximation, to close the system of equations is already 
in the literature\cite{Sacherer:1971_envelope} and in Eq. (\ref{eq:dsqsxdtsq}). 
Moreover our previous work on the forces present in laminar flow Coulomb explosion 
problems\cite{Zerbe:2018_coulomb_dynamics,Zerbe:2019_relativistic_dynamics,Zerbe:2019_laminar_flow}
can be straightforwardly extended to the forces in the sample perspective for particle configurations exhibiting laminar flow.

\section{Conclusion}\label{sec:conclusion}

In this work, we have formalized the sample perspective and compared it to the  
population perspective utilized extensively in the literature.  We showed that these
perspectives give fundamentally different statistical results in at least two cases: 1.) the estimation of the emittance and 2.) the determination of derivatives.  As well as providing new approaches to analysis, the sample perspective provides important new physical insights, and is complementary to the insights gained from the population perspective.  A key distinction between the sample and population perspectives is that the inclusing of collisions is critical to an exact analysis within the population perspective, but the sample perspective provides an exact description from knowledge of the interparticle forces alone.  In this work we concentrated on the sample perspective, however a combined analysis utilizing both the population and sample perspectives is a profitable direction for future work. 
 
In this work we used the sample perspective to provide an exceptionally succinct and general analysis of the evolution of the ensemble statistics of non-interacting particles. 
We showed how emittance arises in the non-interacting theory as a constraint on the waist of 
the distribution when focussed.  Specifically, we showed how emittance is conserved in such non-interacting models under the assumption that all particles have the same Lorentz factor, i.e. the same energy, but we point out that emittance changes when the ensemble has a non-zero energy spread as can be calculated by Eq. (\ref{eq:non em gen}).  In practice where the energy
spread is small, the resulting emittance change during experimentally relevant times
is usually much smaller than the inherent emittance of the ensemble.  This is why non-interacting bunches are often treated as conserved emittance systems\cite{Sacherer:1971_envelope,Reiser:1994_book}; 
however, this is not strictly true as can again be seen in Eq. (\ref{eq:non em gen}).

Utilizing planar models used extensively in the UEM community to describe the spreading of pancake bunches, we calculated exactly the emittance dynamics for these models, which is an important theoretical result.  We showed that the emittance dynamics in the planar model are purely a consequence of global and stochastic non-linearities, and that the global non-linearities may be obtained from the\textit{a priori} distribution using mean-field theoretical techniques.   While emittance growth due to non-linearities is well appreciated by the community\cite{Buon:1992_phase_space},  we specifically calculated the emittance growth equations for the cold planar symmetric uniform, Gaussian, and quartic  bimodal distributions.  We explained how these equations may provide insight into disorder induced heating, and we emphasized that 
the stochastic portion of emittance evolution warrants further attention.

The analysis for the cold distributions is easily extended to cases that do not start from rest by including a non-zero velocity in Eq. (\ref{eq:kin}), for example an initial chirp.  Of course,
additional expectations are necessary, but the predictions should be similarly accurate.  In such a case, additional powers of time  become important resulting in the emittance becoming non-linear as seen in Fig. \ref{fig:emittance growth}, but the coefficients in front of the powers of time can be calculated if the initial distribution in position and velocity space is assumed. In fact, the reason we examine the coefficient in front of the squared time is that the case examined here has only a delta function in velocity space which results in all coefficients except the projection to 
position-acceleration space being zero. 

While this formulation requires that the initial conditions are known, we point out that in 
Fig. \ref{fig:emittance growth} that drawing the particles from the same initial distribution results in solutions that can also be thought of as an evolving distribution; that is an $N$-particle ensemble is simply a single sample point in $6ND$-phase-space and we are concerned with the evolution of the spread of a statistic from repeated samples from that space.  This is the perspective we used to calculate the mean emittance growth for our three planar symmetric distributions.  In fact, we can describe higher order moments of this distribution using this approach; however, such a description is statistically intricate. 

An important computational result present here is the spread in the emittance due to finite particle number, 
which we find scales as $\frac{1}{\sqrt{M}}$, where $M$ is the number of macro-particles chosen in the simulation.  
Specifically, this portion of the slope for the uniform distributions was found to be 
$\frac{0.38 \pm 0.11}{\sqrt{M}} \frac{mm \cdot mrad}{s}$ as can be seen in 
Fig. \ref{fig:discrete:vs N}.  On the other hand, both the initially Gaussian and quadratic bimodal distributions 
were found to be distributed about $0 ~\pm$ some value over $\sqrt{M}$ instead as seen in 
Figs. \ref{fig:gaussian discrete:vs N} and \ref{fig:bimodal discrete:vs N}. 
These $\frac{1}{\sqrt{M}}$ scalings suggests a statistical origin.  
As Eq. (\ref{eq:m_em}) describes the emittance growth of non-relativistic, cold distributions, 
it is apparent that this variation in slope is due to stochastic factors in the initial position-acceleration space obtained from 
sampling the underlying $6ND$ population distribution.  Specifically, the estimates we obtained for $\epsilon_{x,p}$ from 
mean-field theory lack these stochastic factors which should provide ``error bars'' on our predictions.  
Of course, as the mean-field volume for the uniform real-space distribution is zero, any stochastic fluctuation should 
increase this volume resulting in the expectation that the fluctuations introduce a solely positive slope component consistent 
with $0.38 \frac{mm \cdot mrad}{s}$. 
On the other hand, both the Gaussian and bimodal distribution have significant initial volume,  
so it is no surprise that (a.) this volume dominates the
stochastic factor and (b.) stochastic factors appear to be equally likely to drive this 
volume either up or down consistent with the expectation of $0 \frac{mm \cdot mrad}{s}$
we observed for these cases.    
  
Notice, though, that the $\frac{1}{\sqrt{M}}$ scaling was determined from computations, and a more complete statistical model  should have additional coupling terms between the stochastic effects and the geometric details.  Specifically, one would expect that a distribution closer to the
uniform distribution should have a finite stochastic contribution between $0$ and $0.38 \frac{mm \cdot mrad}{s}$.  In fact, we examined the emittance growth of the initially semi-circular
distribution, not presented, and observed evidence of this behavior, but the means of the scaled remainder slope, $m_{stoch}$, 
for differing values of $M$ were not the same until $M>10,000$ 
suggesting  a more complex general form for the deviation in slope other than $\frac{1}{\sqrt{N}}$, particularly at small $N$.  

We also presented how the covariance statistics relate to kinetic energy, and showed how the kinetic energy can 
be decomposed into a linear flow energy and a linear heat.  
For highly
non-equilibrium situations, temperature does not have a standard definition; however, kinetic energy and rms emittance
do.  As can be seen in Appendix \ref{ap:polytropic}, rms emittance conservation 
is occurs for non-interacting, non-relativistic adiabatic expansion, 
which is common place in accelerator physics.  Furthermore, we showed that 
the mathematical formalism presented in this work can be extended to describe non-interacting relativistic dynamics and 
non-relativistic interacting systems in a 
straightforward manner.  We will show in future work that relativistic interacting dynamics can likewise be explained
using this formulation.  Therefore, this relation between rms emittance and what we termed the linear heat,
which can be thought of as the kinetic energy left over once the
flow energy has been approximated through linear regression, is of critical importance for understanding 
systems in UEM/UED and accelerator physics.

We believe that 
this relation between emittance and linear heat may be
why many scholars have tried a thermodynamic interpretation of emittance; specifically, emittance has long been 
suggested as being related to 
entropy\cite{Lapostolle:1971_envelope_filamentation,Struckmeier:1996_entropy,Brown:1996_entropy,Brown:1997_free_energy,OShea:1996_free_energy,OShea:1998_entropy,Boine:2015_intense_beams}.  However, 
we show in Appendix \ref{ap:polytropic} that  under near equilibrium conditions
the emittance is neither intensive nor extensive; in contrast the entropy is extensive.  This leads us to conclude
that emittance and entropy are not directly related.  Of course, if the adiabatic expansion is reversible,
we'd see that both the  emittance and
entropy are conserved, but this does not mean that emittance and entropy are the same thing.  
One way to see some difference
is to examine the expansion of a spherically symmetric continuous Gaussian distribution. Prior to the emergence
of the shock, which breaks the laminar assumption, such a distribution is in fact reversible; however,
the emittance does in fact grow -- if only from non-linear effects.  

Similar arguments concerning the thermodynamic properties of emittance have previously been presented by 
Bernal\cite{Bernal:2015_conceptual}; further, Bernal concluded that an understanding of the mechanisms involved in 
emittance change was needed to advance beyond the current phenomenological understanding.  
The statistical kinematic presentation of emittance change presented in Eq. (\ref{eq:demdt}) provides such an avenue 
for understanding such mechanisms as we have explicitly demonstrated in the planar symmetric.  In the planar symmetric 
case, only global and stochastic  non-linearities play a role.  Additional mechanisms are present in higher 
dimensional problems, and we will examine such mechanisms in future works.

\section{Data availability}
The data that support the findings of this study are available from the corresponding author upon reasonable request.

\begin{acknowledgments}
This work was supported by NSF Grant numbers RC1803719 and RC108666.  We thank Steve Lund for introducing
us to the work of Sacherer. We thank Omid Zandi for helpful discussion concerning emittance growth relations with fluid techniques 
and entropy.
\end{acknowledgments}

\appendix

\section{Mathematical details of emittance growth calculations}\label{ap:emittance math} 

Denote the $i^{th}$ particles position and velocity at time $t=0$ by by $x_{0,i}$ and $v_{0,i}$, respectively, and
at time $t$ by $x_{i}$ and $v_{i}$, respectively.  In the laminar planar model, the force on the $i^{th}$ particle, $F_i$ is 
constant, so assuming non-relativistic conditions, the $i^{th}$ particle's acceleration, $a_i = \frac{F_i}{m}$, is also a constant.
Thus the kinematic equations for the $i^{th}$ particle are given by Eq. (\ref{eq:1D kin:x}).
Notice that $x_i$, $v_i$, $x_{0,i}$, $v_{0,i}$, and $a_i$ differ among the particles, i.e. they are values of the random variables
$x$, $v$, $x_0$, $v_0$, and $a$.  Using this notation, we can write
\begin{subequations}\label{eq:kin}
  \begin{align}
    x &= x_{0} + v_{0} t + \frac{1}{2} a t^2\\
    v &= v_{0} + a t
  \end{align}
\end{subequations}
to be the map between the random variables using the ensemble constants of $\frac{1}{2}$, $t$, and $t^2$.

The map in Eq. (\ref{eq:kin}) can be used to determine the evolution of any statistic.  For example, the
evolution in the velocity variance can be written as
\begin{align}
  s_{v}^2 &= s_{v,v}\nonumber\\
               &= s_{v_{0} + a t,v_{0} + a t}  
\end{align}
The right hand side of this equation can be simplified using Eq. (\ref{eq:cov sum}) giving
\begin{align}
  s_{v}^2 &= s_{v_{0}}^2 + s_a^2 t^2 + 2 s_{v_0,a} t  
\end{align}
Analogously,
subbing
Eq. (\ref{eq:kin}) into Eq. (\ref{eq:def em}) and simplifying with Eq. (\ref{eq:cov sum}), 
we obtain
\begin{align}
  c^2 \epsilon_{x,p}^2 &=   s_{x_0}^2 s_{v_0}^2 - s_{x_0,v_0}^2 \nonumber\\
                                   &\quad+ 2 (s_{x_0}^2s_{v_0,a} - s_{x_0,v_0}s_{x_0,a})t \nonumber\\
                                   &\quad + (s_{x_0}^2s_{a}^2 - s_{x_0,a}^2 - s_{x_0,a}s_{v_0}^2 + s_{x_0,v_0}s_{v_0,a})t^2\nonumber\\
                                   &\quad + (s_{x_0,v_0}s_{a}^2 - s_{x_0,a}s_{v_0,a})t^3\nonumber\\
                                   &\quad + \frac{1}{4} (s_{v_0}^2s_{a}^2 - s_{v_0,a}^2)t^4\label{eq:full 1D em}
\end{align}
The right hand side of this equation can be shown to be equivalent to the determinant of Eq. 
(\ref{eq:time dep matrix}).

\section{Emittance of a ideal monatomic gas in a cubic box at equilibrium}\label{ap:polytropic}
We first consider the emittance of an ideal gas in a box whose edges all have width $\Delta x$
and whose volume is $V = (\Delta x)^3$.  
We assume that the gas is at equilibrium 
meaning that there is no correlation between any of the 6D dimensions,
$x$, $y$, $z$, $p_x$, $p_y$, and $p_z$, 
the distribution and that the spatial dimensions are distributed uniformly,
and the momentum distribution follows the Maxwell-Boltzmann distribution.  
Using the definition of emittance in Eq. (\ref{eq:sample em}),
we obtain under these conditions
\begin{align}
  \varepsilon_{x,p_x}^2 &= \frac{1}{m^2c^2}s_x^2 s_{p_x}^2
\end{align}
If the three spatial dimensions, $x$, $y$, and $z$ are equivalent as they are in a cubic box,
$\frac{s_{p_x}^2}{2 m} = \frac{k_B}{2} T$.  Thus we may write 
\begin{align}
  s_{p_x}^2 &= m k_B T
\end{align}
in this case.  Likewise, $s_x^2$ may be related to the volume, $V$, of a closed container;
the exact expression depends on the geometry of the distribution.  In a cubic box, $V = \Delta x^3$
with a uniformly distributed gas, $s_x = \frac{1}{\sqrt{12}} \Delta x$.  Thus
\begin{align}
  s_x^2 &= \frac{V^{2/3}}{12}
\end{align}
Thus, the emittance can be written as
\begin{align}
  \varepsilon_{x,p_x}^2 &=  \frac{1}{12} \frac{k_B T}{mc^2} V^{2/3}
\end{align}
So, we see that emittance is proportional to the product  volume raised
to the $2/3$ power and temperature in a thermalized cubic box. 

In the language of thermodynamics, temperature is intensive while volume is extensive,
so emittance is neither intensive nor extensive, but a combination of intensive and extensive 
properties.  In contrast, entropy is extensive.
This has been pointed out previously by Bernal in his excellent paper arguing why
the free energy model of emittance growth is problematic\cite{Bernal:2015_conceptual}.  
However, to students of thermodynamics,
the product $T V^{2/3}$ should be familiar for another reason --- 
it is the conserved quantity during the adiabatic expansion of an ideal monoatomic gas
under non-relativistic conditions. 

\nocite{*}
\bibliographystyle{apsrev4-1}
\bibliography{sample_picture_1d}

\end{document}